\documentclass[aps,prl,twocolumn,superscriptaddress]{revtex4-1}
\usepackage{amssymb}
\usepackage{physics}
\usepackage{graphicx}
\usepackage{amsmath}
\usepackage{amsthm}
\usepackage{amsfonts}
\usepackage[T1]{fontenc}
\usepackage{array}
\usepackage{multirow}
\usepackage{color}
\usepackage{esint}
\usepackage{bm}
\usepackage{color}
\usepackage{bbm}
\usepackage{hyperref}
\usepackage{babel}
\usepackage{titlesec}

\newcommand{\AM}{A} 
\newcommand{\lettersection}[1]{\emph{#1}.---}

\usepackage{hyperref}
\hypersetup
{	colorlinks,%
	citecolor=green,%
	linkcolor=red,%
	urlcolor=blue%
}
\allowdisplaybreaks[4]
\begin{document}

\global\long\def\id{\mathbbm{1}}
\global\long\def\ui{\mathbbm{i}}
\global\long\def\ud{\mathrm{d}}

\title{Fermionic Many-Body Localization for Random and Quasiperiodic Systems 
in the presence of Short- and Long-Range Interactions}

\author{DinhDuy Vu}
\affiliation{Condensed Matter Theory Center and Joint Quantum Institute, University of Maryland, College Park, Maryland 20742, USA}
\author{Ke Huang}
\affiliation{Department of Physics, City University of Hong Kong, Kowloon, Hong Kong SAR}
\author{Xiao Li}
\affiliation{Department of Physics, City University of Hong Kong, Kowloon, Hong Kong SAR}
\affiliation{City University of Hong Kong Shenzhen Research Institute, Shenzhen 518057, Guangdong, China}
\author{S. Das Sarma}
\affiliation{Condensed Matter Theory Center and Joint Quantum Institute, University of Maryland, College Park, Maryland 20742, USA}

\begin{abstract}
We study many-body localization (MBL) for interacting one-dimensional lattice fermions in random (Anderson) and quasiperiodic (Aubry-Andre) models, focusing on the role of interaction range.  
We obtain the MBL quantum phase diagrams by calculating the experimentally relevant inverse participation ratio (IPR) at half-filling using exact diagonalization methods and extrapolating to the infinite system size. 
For short-range interactions, our results produce in the phase diagram a qualitative symmetry between weak and strong interaction limits. 
For long-range interactions, no such symmetry exists as the strongly interacting system is always many-body localized, independent of the effective disorder strength, and the system is analogous to a pinned Wigner crystal. 
We obtain various scaling exponents for the IPR, suggesting conditions for different MBL regimes arising from interaction effects. 
\end{abstract}

\maketitle

\lettersection{Introduction}
Many-body localization (MBL), an extensively recently studied phenomenon in quantum statistical mechanics~\cite{ Nandkishore2015_Review,Altman2015_Review,Abanin2019_RMP,Gopalakrishnan2020_Review}, deals with the important subject of thermalization in isolated disordered interacting quantum systems, where all eigenstates (i.e. the whole many-body spectrum, not just the ground state) may become localized for strong enough disorder even in the presence of interactions, thus preventing the system from achieving equilibrium although interaction among the particles should allow energy transport, in principle. 

MBL, which is a generalization of the well-known ground-state Anderson localization~\cite{Anderson1958} to the whole many-body interacting spectrum, is a counterexample to the quantum eigenstate thermalization hypothesis~\cite{Deutsch1991,Srednicki1994}, and as such, violates the basic premise of quantum statistical mechanics. 
Little is known about MBL theoretically, as the standard analytical tools of theoretical physics do not quite apply in dealing with the dynamics of interacting disordered systems in the nonequilibrium thermodynamic limit. 
However, an early work using perturbation theory on a Bethe lattice hinted at the existence of the MBL transition from extended thermal to localized nonthermal situation~\cite{Basko2006}, and a rigorous argument exists for the existence of a localized spectrum (but, not for the MBL transition) in interacting spin chains for sufficiently large disorder~\cite{Imbrie2016,Imbre2016}. 
Most theoretical works on MBL therefore resort to small-system numerics on one-dimensional lattices, which clearly shows the existence of MBL in both purely random Anderson-type (A) disorder model and quasiperiodic Aubry-Andre-type (AA) quasiperiodic model~\cite{AAModel_1980}, at least for finite systems~\cite{Oganesyan2007_PRB,Znidaric2008,Pal2010,Devakul2015,Li2015_PRL,Hsu2018_PRL}. 
In the current work we use exact diagonalization methods to study fermionic MBL in {\AM} and AA models, both for short-range (SR) and long-range (LR) interactions. 
Our physically motivated fermionic long-range interaction is qualitatively different from, but nevertheless, can be mapped to an anisotropic spin model with the long-range $ZZ$ coupling.
In contrast, the long-range $XX$ and $YY$ components do not have a physical fermionic counterpart because they intrinsically involve long-range hoppings, which are always suppressed exponentially in physical systems. 
In principle, long-range hopping can cause delocalization as discussed in Refs.~\cite{Khatami2012,Singh2017,Giuseppe2019,Thomas2019,Yao2014,Burin2015,Gutman2016}. 
This qualitatively distinguishes our study from previous works on spin models~\cite{Wu2016,Yao2014,Burin2015,Gutman2016}.
We also note that fermionic MBL with long-range interactions has been discussed in the context of Luttinger liquids~\cite{Nandkishore_2017,Akhtar_2018} and self-consistent mean field theory~\cite{Roy2019}. These treatments apply to at most moderate interaction, comparable to the energy scale of the non-interacting counterpart, but not to very strong interacting regime where the Hilbert space is immensely partitioned. Our study indeed shows that MBL persists longer for long-range interactions with weak to moderate strength, consistent with other theoretical approaches. However, as shown in our phase diagrams (see the $U\le 5$ regime of Fig.~\ref{Fig3}), this effect is mostly quantitative compared to the much more apparent interaction-range dependence observed in the strongly interacting regime. This is the key advancement in our work.
Finally, MBL has been reported in experimental studies of ultracold atoms on lattices~\cite{Schreiber2015_Science,Lueschen2017_PRL,Lueschen2018_PRL,Kohlert2019_PRL}, where our work applies directly. 
 
In the current work we aim at the ambitious challenge of calculating the MBL quantum phase diagram in the disorder-interaction space for interacting fermions on a (half-filled) 1D lattice, separately considering both short-range and long-range inter-particle interactions. 
The key new features of our work are: (1) the consideration of the whole range of interaction, from weak to strong, in obtaining the MBL phase diagram, and (2) the consideration of LR interaction (in addition to the SR interaction, used extensively in the existing MBL literature). 
We find that for strong LR interactions, MBL exists generically in both {\AM}/AA models for any disorder, no matter how small the disorder is.  
For SR interactions, we find an approximate symmetry between MBL properties at very weak and very strong interactions, which arises from the nature of the Hubbard model employed in our work. 
For weak to modest interaction, SR/LR interactions produce similar results even though, quantitatively, the MBL phase slightly extends for LR interaction. 
These two cases studied here can help us understand the localization property of a generic interaction case.

\lettersection{Model}
The physical systems in our paper are modeled by the following Hamiltonian of spinless electrons: 
\begin{equation}
 H = \sum_i \qty[\qty(c_i^\dagger c_{i+1} + \text{h.c.}) + V_i n_i] +\dfrac{1}{2}\sum_{i,j} U_{i,j}n_in_j.
 \label{eq:Hamiltonian}
\end{equation}
Here the AA potential is $V_i=V\cos (2\pi i /\beta + \phi)$ with $\beta=(\sqrt{5}+1)/2$ and a random phase $\phi$, while for the random A potential, each $V_i$ is picked from a uniform distribution $\qty[-V,V]$. 
The hopping is limited to nearest sites, reflecting the realistic situation in 1D fermionic systems including both solid state lattices  and cold-atom arrays. The effect of electron-electron interaction is studied in two representative cases: the SR nearest-neighbor interaction $U_{i,j}=U\delta_{i\pm 1, j}$, and the LR interaction
\begin{align}\label{eq:interaction}
	U_{i,j} = U\qty(\dfrac{L}{\pi}\sin\dfrac{\pi\abs{i-j}}{L})^{-\kappa}. 
\end{align}
For now we only consider $\kappa = 1$. 
Together, we present four 2D phase diagrams in the interaction-disorder plane: AA model (SR), AA model (LR), {\AM} model (SR) and {\AM} model (LR). 
Our goal is the localization/extension property in the thermodynamic limit. 
We expect that a qualitative picture can be systematically obtained from finite-size calculations.
In particular, for each type of disorder and interaction, we exactly diagonalize finite-size Hamiltonians of different numbers of sites $L$ and electrons $N_e$ (with the filling fraction $\nu=N_e/L$ fixed at $1/2$). 
The results are then extrapolated to the limit $L\to \infty$, providing predictions about the half-filled system in the thermodynamic limit~\cite{Supplement, Khatami2012, Nag2017}. We impose the periodic boundary condition $c_{L+1}=c_1$ to eliminate the boundary effects in finite-size calculations.

After diagonalizing the finite-size Hamiltonian~\eqref{eq:Hamiltonian}, the localization property of each eigenstate $\psi_{\lambda}$ ($\lambda$ is the label of many-body eigenstates) can be quantified by the scaled (inverse participation ratio) IPR \cite{Bera2015}, defined as 
\begin{equation}
	\mathcal{I}^{(\lambda)} = \dfrac{1}{1-\nu}\qty(N_e^{-1}\sum_{i=1}^L \abs{u_i^{(\lambda)}}^2-\nu), \label{Eq:IPR}
\end{equation}
where $u_i^{(\lambda)} \equiv \expval{n_i}{\psi_{\lambda}}$ is the projected particle number of each state ${\psi_{\lambda}}$ on site $i$. 
In the maximally extended state, $u_i=N_e/L$, resulting in $\mathcal{I}=0$; while in the opposite limit, $u_i=1$ ($0$) for an occupied (vacant) site, corresponding to $\mathcal{I}=1$. 
Therefore, a generic eigenstate has $0\le \mathcal{I} \le 1$. 
To benchmark the degree of localization of the whole spectrum, we use the infinite-temperature (arithmetic mean) scaled IPR, $\expval{\mathcal{I}} \equiv \sum_\lambda \mathcal{I}^{(\lambda)}/d$, with $d$ being the dimension of the Hilbert space. 
For brevity, we will refer to $\expval{\mathcal{I}}$ simply as the mean IPR.

\begin{figure}
	\includegraphics[width=0.48\textwidth]{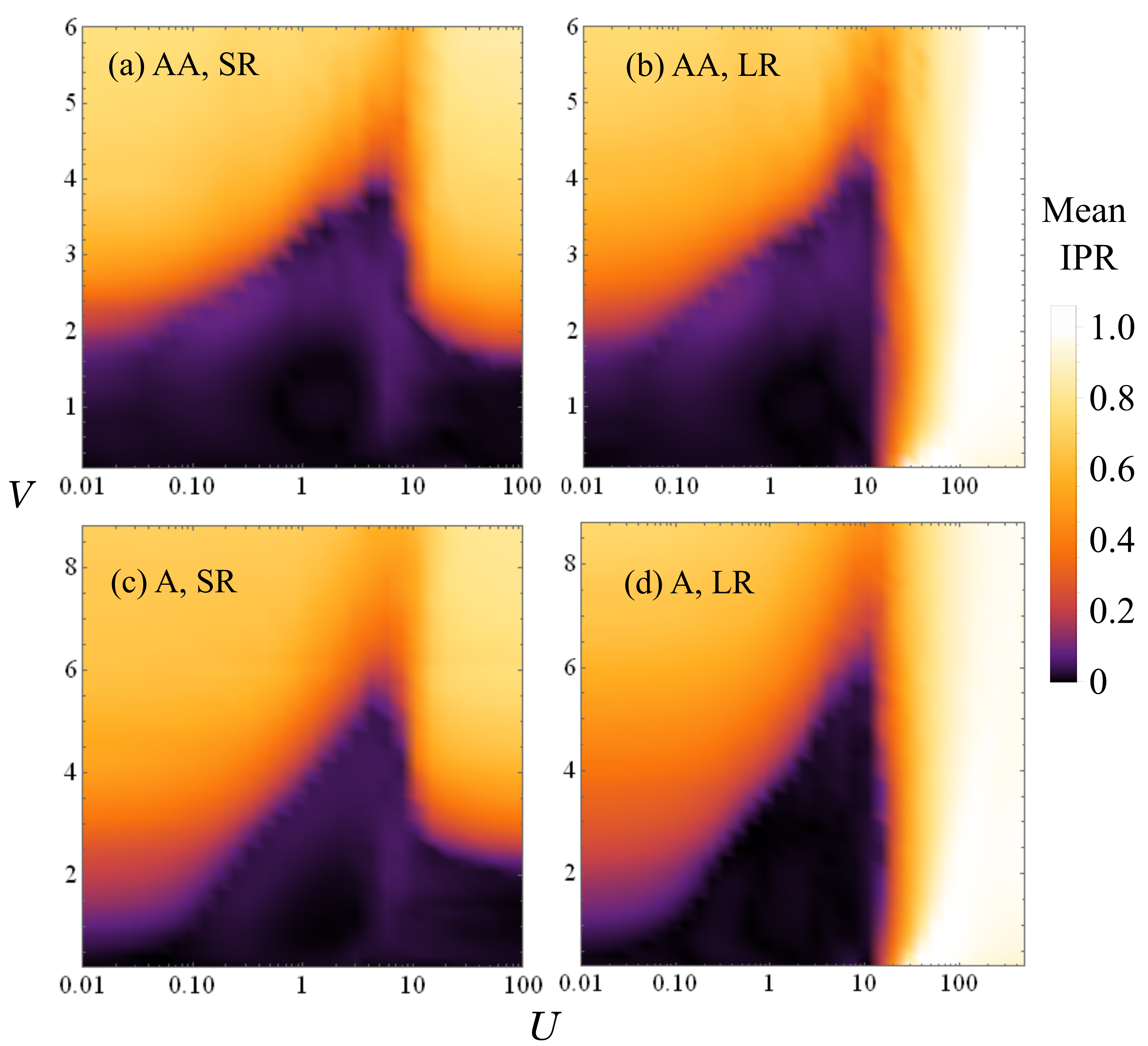}
	\caption{Extrapolated quantum phase diagram from simulations on finite-size lattices of $L=8,10,\dots,16$. 
\label{Fig1}}
\end{figure} 

\lettersection{Quantum phase diagram} 
By exact diagonalization, we compute the mean IPR as a function of $U$ and $V$ for five different system sizes, $L=8,\cdots,16$ at half filling, and extrapolate these results into the limit $N_e\to \infty$ to obtain the phase diagram. 
The AA model results are averaged over six random phases, while the {\AM} model results are averaged over six random disorder realizations. This number of realizations is sufficient for our purpose, as confirmed by our numerical investigation.
The calculated MBL phase diagrams are presented in Fig.~\ref{Fig1}.

\begin{figure}
\includegraphics[width=0.45\textwidth]{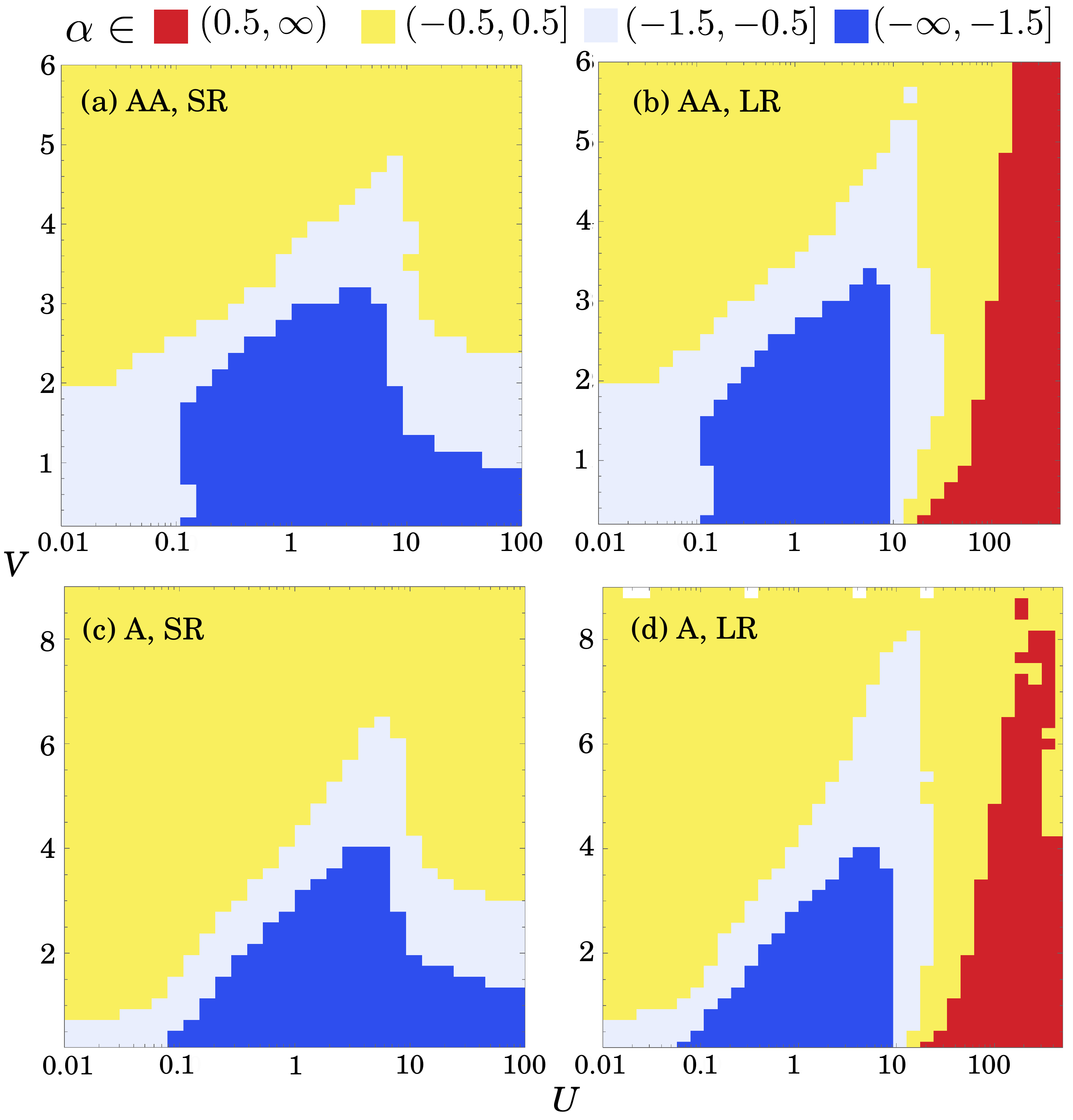}
\caption{The effective scaling exponent $\alpha$ is divided into four color-coded regimes: 
(i) the marginal localized phase $\alpha\approx 0$ (yellow), 
(ii) the reciprocal extended phase $\alpha\approx-1$ (light blue), 
(iii) the relevant localized phase $\alpha>0$ (red), and 
(iv) the super-reciprocal extended phase $\alpha < -1$ (dark blue). \label{Fig2}} 
\end{figure}

It is well established that the noninteracting AA model has a critical point at $V=2$, with the whole spectrum localized (extended) for $V > 2$ ($V<2$). 
By contrast, the {\AM} model is always localized for any $V>0$ (we refer to \cite{Supplement} or Figs.~\ref{Fig2}(c) and (d) for better visibility).
Figure~\ref{Fig1} shows that, for both models of interactions, this noninteracting localization persists in the regime $U\ll 1$, consistent with the statement that MBL exists for perturbatively weak interactions~\cite{Basko2006,Roy2019,Sabyasachi2019}. This might also apply to some cases of systems with mobility edge \cite{Nag2017}.
However, when $U\sim\mathcal{O}(1)$, there exists a transition from the MBL to the thermal phase (as a function of $U$) that depends highly on the underlying non-interacting model but quantitatively little on the interaction type.  
Specifically, in the extrapolated thermodynamic limit, the thermal phase extends to $V\sim 4$ for the AA model and $V\sim 6$ for the {\AM} model, above which the corresponding interacting phase is always MBL. 
This transition might be understood from the reordering of Fock state energy levels (disorder plus interaction potential), resulting in level crossings as $U$ increases. Around such crossings, the Fock basis states can be strongly hybridized by the kinetic energy term. Even though these crossing points do not necessarily happen at the same $U$, their resonant regions, where the Fock basis hybridization is substantial, might be overlapping for sufficiently large electron hopping, leading to the delocalization of the whole spectrum. Importantly, the above argument becomes invalid if $V,U \gg 1$ (the weak hopping limit) and the resonant regimes are diluted, rendering MBL generic in the very weak hopping limit.  A mathematically rigorous proof can be found in Refs.~\cite{Imbrie2016,Imbre2016,Mastropietro2015} for sufficiently large disorder, but it appears that this limit is reached already for $V\sim4$--$6$ in Fig.~\ref{Fig1}.  We cannot rule out the possibility that the critical MBL disorder depends somewhat on the system size \cite{Devakul2015}. 
 
Up to the intermediate region $U\sim \mathcal{O}(1)$, we find no qualitative differences between LR and SR results. 
However, this is no longer true for $U\gg1$.
Particularly, the strongly interacting phase with LR interaction is localized for all value of $V$ for both models of disorder [see Figs.~\ref{Fig1}(b) and (d)]. 
This suggests that for LR interactions, the nature of the strongly interacting regime is disconnected from the noninteracting regime. 
In fact, the localization of all eigenstates when $U\to\infty$ is entirely driven by interactions, which is reminiscent of the mechanism of a Wigner crystal where the role of disorder is to only break the translational symmetry to pin down the Wigner crystal. 
The picture is completely different for SR interactions, as there apparently exist both localized and extended phases in the large-$U$ limit [see Figs.~\ref{Fig1}(a) and (c)]. 
Obviously, disorder must play an important role in the phase transition for SR interactions. We emphasize that in Fig.~\ref{Fig1} both A and AA models behave qualitatively similarly vis-a-vis SR and LR interaction dependence.

To understand the nature of various regions in the phase diagram, we perform a scaling analysis using the ansatz $\expval{\mathcal{I}}/(1-\expval{\mathcal{I}}) \propto N_e^\alpha$ (or equivalently $L^\alpha$ as $L/N_e=2$) \cite{Supplement} and find that the phase diagrams can be partitioned, depending on the value of $\alpha$, into four zones: 
(i) marginal localized ($\alpha\sim 0$), (ii) reciprocal extended ($\alpha\sim -1$), (iii) relevant localized ($\alpha>0$), and (iv) super-reciprocal extended ($\alpha<-1$).
The results are presented in Fig.~\ref{Fig2}. 
One can think of the system size scaling process as gluing identical lattice segments, thus the scaling behavior depends on how the bulk ``communicates'' with the boundary. 
The disorder-driven localized phase obviously does not know about the boundary, leading to the marginal scaling ($\alpha=0$). This is the standard MBL phase mostly discussed in the existing literature.
On the other hand, for the extended or thermal phase, all electronic states can spread through the boundary to the extension lattice, suppressing the density fluctuations. 
Specifically, the density profile of a noninteracting many-body eigenstate is simply the sum of each constituent single-particle eigenstate, $n_\text{MB}(i) = \sum_j n_{\beta_j}(i)$, where MB stands for ``many-body'',  $i$ labels the lattice sites, and $\beta_j$ is the single-particle eigenstate index with $j=1,\dots,N_e$. 
Since all single-particle states are extended, we assume the form $n_{\beta_j}(i)=[1+\phi_{\beta_j}(i)]/L$, where $\phi_{\beta_j}(i)$ is the oscillating part integrated to zero, analogous to the familiar Friedel oscillations. Because of the oscillating nature,  $\sum_{i,j,k}\phi_{\beta_j}(i)\phi_{\beta_k}(i)$ must be subleading compared to $L^3$, thus
\begin{align}
	\sum_{i=1}^{L} n_\text{MB}(i)^2 = \nu N_e + \mathcal{O}(1). 
\end{align}
If we then replace $\sum_i^{L} \abs{u_i^{(\alpha)}}^2$ in Eq.~\eqref{Eq:IPR} by $\sum_i n_\text{MB}(i)^2$, we find that $\expval{\mathcal{I}}\sim N_e^{-1}$, consistent with the observed reciprocal scaling $\alpha=-1$ if the extended phase is adiabatically connected to the single-particle picture. 

However, the strongly interacting extended phase is more complicated. 
Numerical simulations show that this phase has a stronger decreasing scaling behavior. 
Compared to its noninteracting counterpart, an interacting many-body eigenstate cannot be expressed as a product of single-particle wavefunctions. 
In fact, from the level statistics information, we know that the interacting extended phase follows the Gaussian orthogonal ensemble (GOE)~\cite{Supplement}, i.e., no information of the original noninteracting product states can be retrieved from the interacting correlated states. 
As a result, the density fluctuations of the correlated extended eigenstates have higher order corrections (beside the homogeneous density distribution), leading to the super-reciprocal scaling exponent $\alpha < -1$. 
We do not rule out the possibility that $\expval{\mathcal{I}}$ might decay even faster than a power-law function of $L$. 

\lettersection{Strong-$U$ generic localization for LR interactions}  We now turn to the localized phase with relevant scaling. We emphasize that this phase only appears in the LR interaction case for very large $U$, unambiguously establishing it as a strongly correlated phenomenon. In fact, the change from the extended to the interaction-induced MBL at $U\sim 10$ looks like a sharp phase transition. The absence of this phase in the SR interaction case is also an interesting point and will be discussed later. In the limit $U\to \infty$, only Fock basis states having the same interaction energies can hybridize; 
for the LR interaction, such states can only be related by a global translation operator, i.e. states are related by shifting all occupied sites by the same amount. 
This immediately rules out a direct coupling through a single electron hopping, as this process only moves one occupied site at a time. In fact, we find that the lowest-order intraband coupling must scale as $(1/U)^{N_e}$, while the band splitting only scales polynomially $\sim  N_e^\gamma$. This suppresses the intra-band resonance as the system size grows \cite{Supplement}, resulting in the apparent relevant scaling behavior. 
This scaling clearly establishes that the localization is driven by the interaction. Similar to Wigner crystal, the LR characteristic of the interaction is decisive here \cite{Vu2020}. On the other hand, for SR interactions, the scaling is marginal at large interactions, similar to the noninteracting case, suggesting a qualitatively different physics. 

\begin{figure}
	\includegraphics[width=0.45\textwidth]{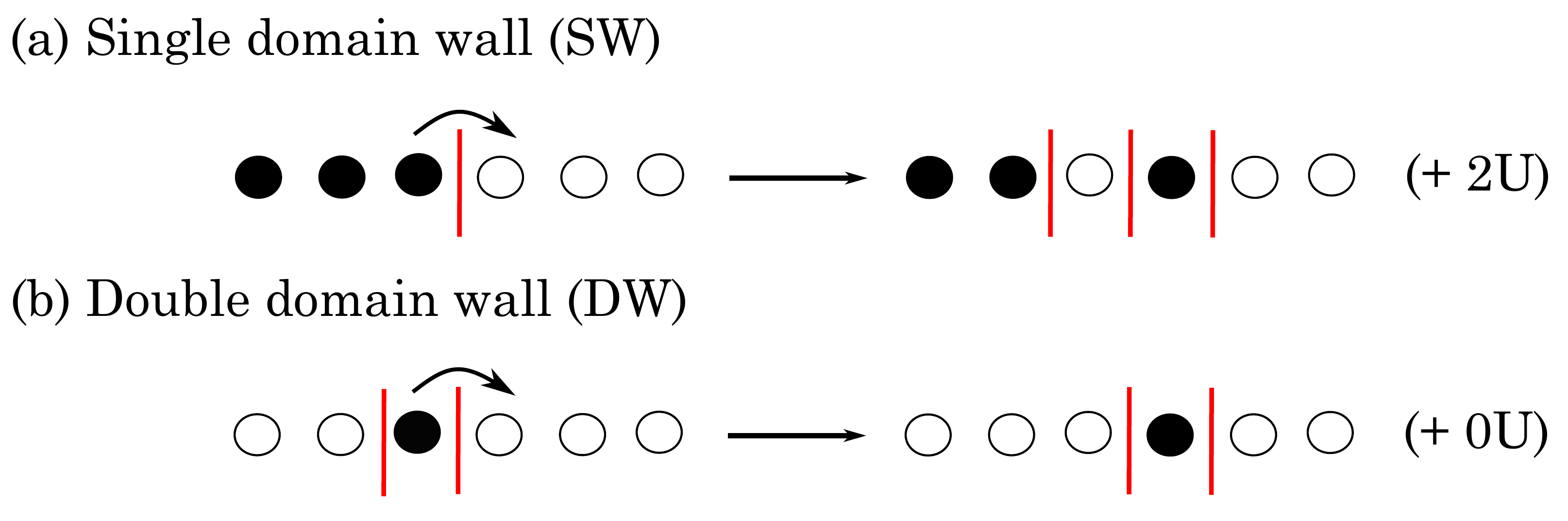}
	\caption{(a) An immobile single domain wall and (b) an itinerant double domain wall. \label{Fig5}}
\end{figure}

\lettersection{Symmetry between strong-$U$ and weak-$U$ limits for SR interactions} 
Compared with LR interactions, the constraints imposed by the SR interaction do not completely exclude direct coupling terms between Fock basis states. 
We can identify an occupied/vacant site with an up/down spin, effectively mapping the electron model to the Heisenberg spin chain where, in the strong-$U$ limit, dynamical processes must conserve the number of domain walls between one occupied and one vacant site. 
We consider two types of domain walls: a single domain wall (SW) with no other domain walls in its immediate vicinity, and double domain walls (DW) formed by two domain walls next to each other. 
Upon applying a single electron hopping to an SW, two additional domain walls are created, thus violating the conservation of domain wall number. 
Therefore, an SW is immobile when $U\to \infty$. 
On the other hand, the same operator moves the DW by one lattice constant without changing the total number of domain walls [see Figs.~\ref{Fig5}(a) and (b)] \cite{DeTomas2019,Langlett2021}. 
This is the intra-band direct coupling that distinguishes between LR and SR cases when $U\to\infty$. 
A DW can thus propagate ballistically throughout the system like a free particle and is also subject to the disordered on-site potential. 
This simple picture suggests the symmetry between the $U\to0$ and $U\to\infty$ limits for SR interactions. \footnote{In particular, the IPR at fixed $V$ as a function of $U$ should demonstrate similar values for $U=0$ and $U\to \infty$. We note that the domain wall argument is exact in the strong-$U$ limit and should be applicable non-perturbatively (in $V$). However, the symmetry is not quantitatively exact for small systems but should improve as the system size grows ~\cite{Supplement}}.

\begin{figure}
\includegraphics[width=0.44\textwidth]{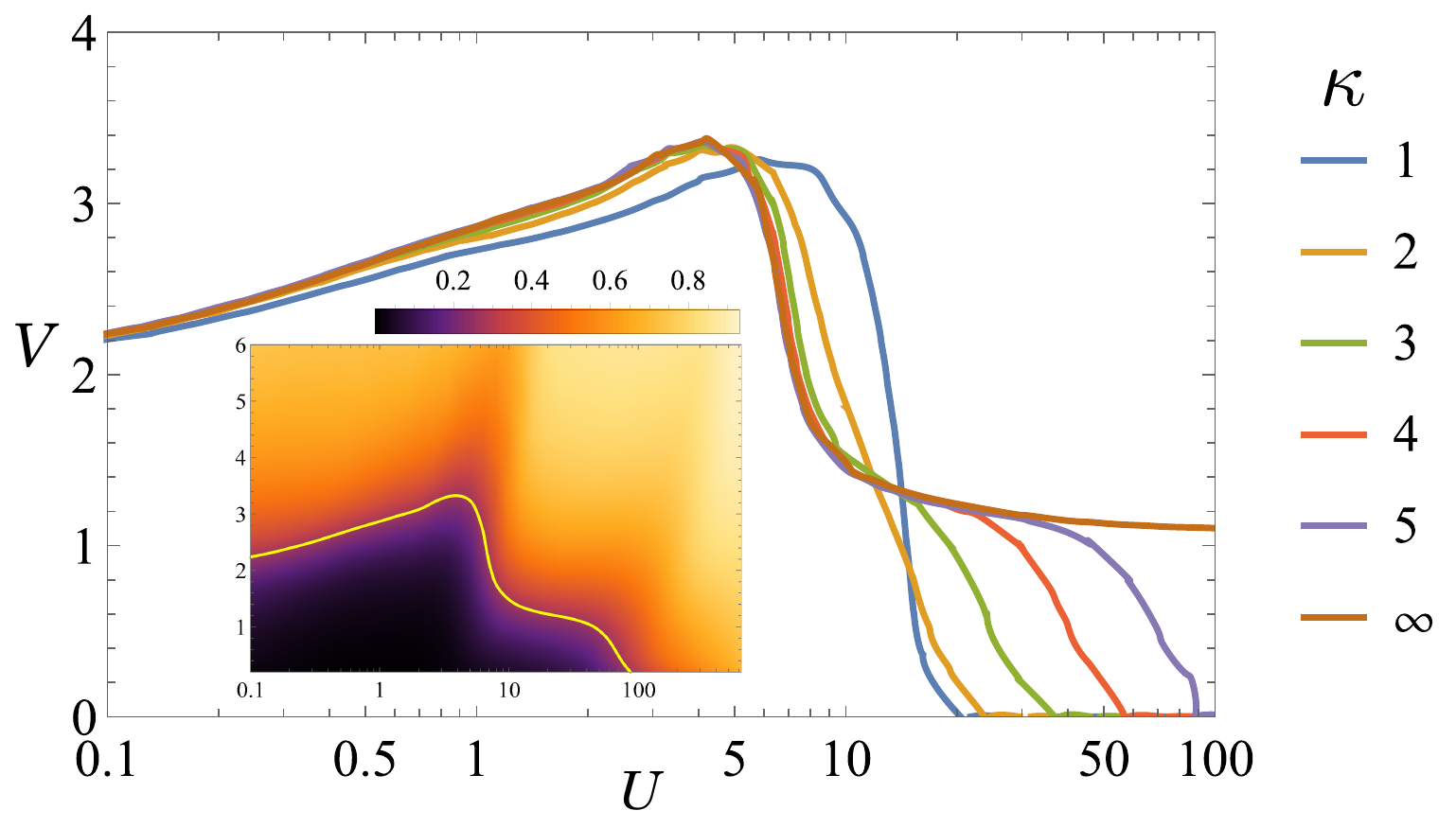}
\caption{\label{Fig3} 
Phase boundaries at different exponent $\kappa$ defined by $\expval{\mathcal{I}}=0.27$ for $L=14$ so that they start at $V=2$ for $U=0$. The inset shows the mean IPR phase diagram for the exponent of $5$ with the previously defined phase boundary. } 
\end{figure}  	


The domain wall argument for SR interaction and the Fock state splitting for LR interaction that we presented above in the strongly interacting regime are actually forms of Hilbert space fragmentation~\cite{DeTomas2019,Pablo2020,Vedika2020,Langlett2021}. 
The striking differences between the LR and SR cases arise because the coupling within each fragment is suppressed by $U$ in the LR case while it is the original electron hopping in the SR case. 
In cold atom experiments, the hopping most likely occurs between nearest neighbors, while the interaction can assume a generic power-law function, e.g. $r^{-3}$ for dipole-dipole interaction. 
Therefore, it is natural to ask whether our results in the strong interaction limit can be generalized to other forms of interactions. 
To this end, we now consider a more general form of interaction with $1 < \kappa < \infty$ in Eq.~\eqref{eq:interaction}. 
The results presented in Fig.~\ref{Fig3} indicate that the phase boundaries with different $\kappa$ for $L=14$ coincide with the SR case ($\kappa\to \infty$) until a critical value $U_{c}^{(\kappa)}$ then eventually reduce to zero, similar to the LR case ($\kappa=1$). 
The SR behavior should dominate when $U_{i,i+1}>U_c>U_{i,i+N_e}$ while the LR behavior starts to manifest when $U_{i,i+N_e}>U_c$ with $U_c$ being some threshold value.
As a result, the critical value $\ln U_{c}^{(\kappa)} \sim \kappa$, which can be seen approximately in Fig.~\ref{Fig3}. 
Therefore, the two cases we study in depth in this work ($\kappa=1$ and $\kappa=\infty$) can be thought of as two limiting cases between which lie other cases of generic interactions. 

\lettersection{Conclusion}
We have presented the MBL quantum phase diagrams for SR and LR interactions in the random A and quasiperiodic AA models over a wide range of disorder and interaction strengths for interacting fermions in 1D tight-binding lattices. We establish that the interaction range plays a nontrivial role in MBL physics.
Our findings include the generic MBL in both A and AA models when the disorder strength is $V>4$--$6$ in lattice hopping units (and there is a generic symmetry between weak and strong interactions in the phase diagram for SR interactions), but MBL is generic for all disorder strengths in the strong LR interaction limit for both A and AA models. The phase diagram of a generic polynomially system should lie between these two limits. 
Experiments using cold atoms on optical lattices can verify our prediction of the generic interaction-driven MBL in LR interacting 1D systems.

\begin{acknowledgments}
\lettersection{Acknowledgements}
This work is supported by Microsoft and Laboratory for Physical Sciences. This work is also generously supported by the High Performance Computing Center (HPCC) at the University of Maryland. 
X.L. also acknowledges support from City University of Hong Kong (Project~No.~11904305), 
the Research Grants Council of Hong Kong (Grants~No.~CityU~21304720 and CityU~11300421), 
as well as the National Natural Science Foundation of China (Grant~No.~9610428).  
\end{acknowledgments}

\bibliographystyle{apsrev4-2}
\bibliography{AA_MBL_v1}

\begin{thebibliography}{44}%
\makeatletter
\providecommand \@ifxundefined [1]{%
 \@ifx{#1\undefined}
}%
\providecommand \@ifnum [1]{%
 \ifnum #1\expandafter \@firstoftwo
 \else \expandafter \@secondoftwo
 \fi
}%
\providecommand \@ifx [1]{%
 \ifx #1\expandafter \@firstoftwo
 \else \expandafter \@secondoftwo
 \fi
}%
\providecommand \natexlab [1]{#1}%
\providecommand \enquote  [1]{``#1''}%
\providecommand \bibnamefont  [1]{#1}%
\providecommand \bibfnamefont [1]{#1}%
\providecommand \citenamefont [1]{#1}%
\providecommand \href@noop [0]{\@secondoftwo}%
\providecommand \href [0]{\begingroup \@sanitize@url \@href}%
\providecommand \@href[1]{\@@startlink{#1}\@@href}%
\providecommand \@@href[1]{\endgroup#1\@@endlink}%
\providecommand \@sanitize@url [0]{\catcode `\\12\catcode `\$12\catcode
  `\&12\catcode `\#12\catcode `\^12\catcode `\_12\catcode `\%12\relax}%
\providecommand \@@startlink[1]{}%
\providecommand \@@endlink[0]{}%
\providecommand \url  [0]{\begingroup\@sanitize@url \@url }%
\providecommand \@url [1]{\endgroup\@href {#1}{\urlprefix }}%
\providecommand \urlprefix  [0]{URL }%
\providecommand \Eprint [0]{\href }%
\providecommand \doibase [0]{https://doi.org/}%
\providecommand \selectlanguage [0]{\@gobble}%
\providecommand \bibinfo  [0]{\@secondoftwo}%
\providecommand \bibfield  [0]{\@secondoftwo}%
\providecommand \translation [1]{[#1]}%
\providecommand \BibitemOpen [0]{}%
\providecommand \bibitemStop [0]{}%
\providecommand \bibitemNoStop [0]{.\EOS\space}%
\providecommand \EOS [0]{\spacefactor3000\relax}%
\providecommand \BibitemShut  [1]{\csname bibitem#1\endcsname}%
\let\auto@bib@innerbib\@empty
\bibitem [{\citenamefont {Nandkishore}\ and\ \citenamefont
  {Huse}(2015)}]{Nandkishore2015_Review}%
  \BibitemOpen
  \bibfield  {author} {\bibinfo {author} {\bibfnamefont {R.}~\bibnamefont
  {Nandkishore}}\ and\ \bibinfo {author} {\bibfnamefont {D.~A.}\ \bibnamefont
  {Huse}},\ }\href {https://doi.org/10.1146/annurev-conmatphys-031214-014726}
  {\bibfield  {journal} {\bibinfo  {journal} {Annu. Rev. Condens. Matter
  Phys.}\ }\textbf {\bibinfo {volume} {6}},\ \bibinfo {pages} {15} (\bibinfo
  {year} {2015})}\BibitemShut {NoStop}%
\bibitem [{\citenamefont {Altman}\ and\ \citenamefont
  {Vosk}(2015)}]{Altman2015_Review}%
  \BibitemOpen
  \bibfield  {author} {\bibinfo {author} {\bibfnamefont {E.}~\bibnamefont
  {Altman}}\ and\ \bibinfo {author} {\bibfnamefont {R.}~\bibnamefont {Vosk}},\
  }\href {https://doi.org/10.1146/annurev-conmatphys-031214-014701} {\bibfield
  {journal} {\bibinfo  {journal} {Annu. Rev. Condens. Matter Phys.}\ }\textbf
  {\bibinfo {volume} {6}},\ \bibinfo {pages} {383} (\bibinfo {year}
  {2015})}\BibitemShut {NoStop}%
\bibitem [{\citenamefont {Abanin}\ \emph {et~al.}(2019)\citenamefont {Abanin},
  \citenamefont {Altman}, \citenamefont {Bloch},\ and\ \citenamefont
  {Serbyn}}]{Abanin2019_RMP}%
  \BibitemOpen
  \bibfield  {author} {\bibinfo {author} {\bibfnamefont {D.~A.}\ \bibnamefont
  {Abanin}}, \bibinfo {author} {\bibfnamefont {E.}~\bibnamefont {Altman}},
  \bibinfo {author} {\bibfnamefont {I.}~\bibnamefont {Bloch}},\ and\ \bibinfo
  {author} {\bibfnamefont {M.}~\bibnamefont {Serbyn}},\ }\href
  {https://doi.org/10.1103/revmodphys.91.021001} {\bibfield  {journal}
  {\bibinfo  {journal} {Rev. Mod. Phys.}\ }\textbf {\bibinfo {volume} {91}},\
  \bibinfo {pages} {021001} (\bibinfo {year} {2019})}\BibitemShut {NoStop}%
\bibitem [{\citenamefont {Gopalakrishnan}\ and\ \citenamefont
  {Parameswaran}(2020)}]{Gopalakrishnan2020_Review}%
  \BibitemOpen
  \bibfield  {author} {\bibinfo {author} {\bibfnamefont {S.}~\bibnamefont
  {Gopalakrishnan}}\ and\ \bibinfo {author} {\bibfnamefont {S.}~\bibnamefont
  {Parameswaran}},\ }\href {https://doi.org/10.1016/j.physrep.2020.03.003}
  {\bibfield  {journal} {\bibinfo  {journal} {Phys. Rep.}\ }\textbf {\bibinfo
  {volume} {862}},\ \bibinfo {pages} {1} (\bibinfo {year} {2020})}\BibitemShut
  {NoStop}%
\bibitem [{\citenamefont {Anderson}(1958)}]{Anderson1958}%
  \BibitemOpen
  \bibfield  {author} {\bibinfo {author} {\bibfnamefont {P.~W.}\ \bibnamefont
  {Anderson}},\ }\href {https://doi.org/10.1103/physrev.109.1492} {\bibfield
  {journal} {\bibinfo  {journal} {Phys. Rev.}\ }\textbf {\bibinfo {volume}
  {109}},\ \bibinfo {pages} {1492} (\bibinfo {year} {1958})}\BibitemShut
  {NoStop}%
\bibitem [{\citenamefont {Deutsch}(1991)}]{Deutsch1991}%
  \BibitemOpen
  \bibfield  {author} {\bibinfo {author} {\bibfnamefont {J.~M.}\ \bibnamefont
  {Deutsch}},\ }\href {https://doi.org/10.1103/PhysRevA.43.2046} {\bibfield
  {journal} {\bibinfo  {journal} {Phys. Rev. A}\ }\textbf {\bibinfo {volume}
  {43}},\ \bibinfo {pages} {2046} (\bibinfo {year} {1991})}\BibitemShut
  {NoStop}%
\bibitem [{\citenamefont {Srednicki}(1994)}]{Srednicki1994}%
  \BibitemOpen
  \bibfield  {author} {\bibinfo {author} {\bibfnamefont {M.}~\bibnamefont
  {Srednicki}},\ }\href {https://doi.org/10.1103/PhysRevE.50.888} {\bibfield
  {journal} {\bibinfo  {journal} {Phys. Rev. E}\ }\textbf {\bibinfo {volume}
  {50}},\ \bibinfo {pages} {888} (\bibinfo {year} {1994})}\BibitemShut
  {NoStop}%
\bibitem [{\citenamefont {Basko}\ \emph {et~al.}(2006)\citenamefont {Basko},
  \citenamefont {Aleiner},\ and\ \citenamefont {Altshuler}}]{Basko2006}%
  \BibitemOpen
  \bibfield  {author} {\bibinfo {author} {\bibfnamefont {D.}~\bibnamefont
  {Basko}}, \bibinfo {author} {\bibfnamefont {I.}~\bibnamefont {Aleiner}},\
  and\ \bibinfo {author} {\bibfnamefont {B.}~\bibnamefont {Altshuler}},\ }\href
  {https://doi.org/10.1016/j.aop.2005.11.014} {\bibfield  {journal} {\bibinfo
  {journal} {Ann. Phys.}\ }\textbf {\bibinfo {volume} {321}},\ \bibinfo {pages}
  {1126} (\bibinfo {year} {2006})}\BibitemShut {NoStop}%
\bibitem [{\citenamefont {Imbrie}(2016{\natexlab{a}})}]{Imbrie2016}%
  \BibitemOpen
  \bibfield  {author} {\bibinfo {author} {\bibfnamefont {J.~Z.}\ \bibnamefont
  {Imbrie}},\ }\href {https://doi.org/10.1007/s10955-016-1508-x} {\bibfield
  {journal} {\bibinfo  {journal} {J. Stat. Phys.}\ }\textbf {\bibinfo {volume}
  {163}},\ \bibinfo {pages} {998} (\bibinfo {year}
  {2016}{\natexlab{a}})}\BibitemShut {NoStop}%
\bibitem [{\citenamefont {Imbrie}(2016{\natexlab{b}})}]{Imbre2016}%
  \BibitemOpen
  \bibfield  {author} {\bibinfo {author} {\bibfnamefont {J.~Z.}\ \bibnamefont
  {Imbrie}},\ }\href {https://doi.org/10.1103/PhysRevLett.117.027201}
  {\bibfield  {journal} {\bibinfo  {journal} {Phys. Rev. Lett.}\ }\textbf
  {\bibinfo {volume} {117}},\ \bibinfo {pages} {027201} (\bibinfo {year}
  {2016}{\natexlab{b}})}\BibitemShut {NoStop}%
\bibitem [{\citenamefont {Aubry}\ and\ \citenamefont
  {Andr{\'e}}(1980)}]{AAModel_1980}%
  \BibitemOpen
  \bibfield  {author} {\bibinfo {author} {\bibfnamefont {S.}~\bibnamefont
  {Aubry}}\ and\ \bibinfo {author} {\bibfnamefont {G.}~\bibnamefont
  {Andr{\'e}}},\ }\href@noop {} {\bibfield  {journal} {\bibinfo  {journal}
  {Ann. Israel Phys. Soc}\ }\textbf {\bibinfo {volume} {3}},\ \bibinfo {pages}
  {18} (\bibinfo {year} {1980})}\BibitemShut {NoStop}%
\bibitem [{\citenamefont {Oganesyan}\ and\ \citenamefont
  {Huse}(2007)}]{Oganesyan2007_PRB}%
  \BibitemOpen
  \bibfield  {author} {\bibinfo {author} {\bibfnamefont {V.}~\bibnamefont
  {Oganesyan}}\ and\ \bibinfo {author} {\bibfnamefont {D.~A.}\ \bibnamefont
  {Huse}},\ }\href {https://doi.org/10.1103/physrevb.75.155111} {\bibfield
  {journal} {\bibinfo  {journal} {Phys. Rev. B}\ }\textbf {\bibinfo {volume}
  {75}},\ \bibinfo {pages} {155111} (\bibinfo {year} {2007})}\BibitemShut
  {NoStop}%
\bibitem [{\citenamefont {{\v{Z}}nidari{\v{c}}}\ \emph
  {et~al.}(2008)\citenamefont {{\v{Z}}nidari{\v{c}}}, \citenamefont {Prosen},\
  and\ \citenamefont {Prelov{\v{s}}ek}}]{Znidaric2008}%
  \BibitemOpen
  \bibfield  {author} {\bibinfo {author} {\bibfnamefont {M.}~\bibnamefont
  {{\v{Z}}nidari{\v{c}}}}, \bibinfo {author} {\bibfnamefont {T.}~\bibnamefont
  {Prosen}},\ and\ \bibinfo {author} {\bibfnamefont {P.}~\bibnamefont
  {Prelov{\v{s}}ek}},\ }\href {https://doi.org/10.1103/physrevb.77.064426}
  {\bibfield  {journal} {\bibinfo  {journal} {Phys. Rev. B}\ }\textbf {\bibinfo
  {volume} {77}},\ \bibinfo {pages} {064426} (\bibinfo {year}
  {2008})}\BibitemShut {NoStop}%
\bibitem [{\citenamefont {Pal}\ and\ \citenamefont {Huse}(2010)}]{Pal2010}%
  \BibitemOpen
  \bibfield  {author} {\bibinfo {author} {\bibfnamefont {A.}~\bibnamefont
  {Pal}}\ and\ \bibinfo {author} {\bibfnamefont {D.~A.}\ \bibnamefont {Huse}},\
  }\href {https://doi.org/10.1103/PhysRevB.82.174411} {\bibfield  {journal}
  {\bibinfo  {journal} {Phys. Rev. B}\ }\textbf {\bibinfo {volume} {82}},\
  \bibinfo {pages} {174411} (\bibinfo {year} {2010})}\BibitemShut {NoStop}%
\bibitem [{\citenamefont {Devakul}\ and\ \citenamefont
  {Singh}(2015)}]{Devakul2015}%
  \BibitemOpen
  \bibfield  {author} {\bibinfo {author} {\bibfnamefont {T.}~\bibnamefont
  {Devakul}}\ and\ \bibinfo {author} {\bibfnamefont {R.~R. P.~P.}\ \bibnamefont
  {Singh}},\ }\href {https://doi.org/10.1103/physrevlett.115.187201} {\bibfield
   {journal} {\bibinfo  {journal} {Phys. Rev. Lett.}\ }\textbf {\bibinfo
  {volume} {115}},\ \bibinfo {pages} {187201} (\bibinfo {year}
  {2015})}\BibitemShut {NoStop}%
\bibitem [{\citenamefont {Li}\ \emph {et~al.}(2015)\citenamefont {Li},
  \citenamefont {Ganeshan}, \citenamefont {Pixley},\ and\ \citenamefont {{Das
  Sarma}}}]{Li2015_PRL}%
  \BibitemOpen
  \bibfield  {author} {\bibinfo {author} {\bibfnamefont {X.-P.}\ \bibnamefont
  {Li}}, \bibinfo {author} {\bibfnamefont {S.}~\bibnamefont {Ganeshan}},
  \bibinfo {author} {\bibfnamefont {J.~H.}\ \bibnamefont {Pixley}},\ and\
  \bibinfo {author} {\bibfnamefont {S.}~\bibnamefont {{Das Sarma}}},\ }\href
  {https://doi.org/10.1103/physrevlett.115.186601} {\bibfield  {journal}
  {\bibinfo  {journal} {Phys. Rev. Lett.}\ }\textbf {\bibinfo {volume} {115}},\
  \bibinfo {pages} {186601} (\bibinfo {year} {2015})}\BibitemShut {NoStop}%
\bibitem [{\citenamefont {Hsu}\ \emph {et~al.}(2018)\citenamefont {Hsu},
  \citenamefont {Li}, \citenamefont {Deng},\ and\ \citenamefont {{Das
  Sarma}}}]{Hsu2018_PRL}%
  \BibitemOpen
  \bibfield  {author} {\bibinfo {author} {\bibfnamefont {Y.-T.}\ \bibnamefont
  {Hsu}}, \bibinfo {author} {\bibfnamefont {X.}~\bibnamefont {Li}}, \bibinfo
  {author} {\bibfnamefont {D.-L.}\ \bibnamefont {Deng}},\ and\ \bibinfo
  {author} {\bibfnamefont {S.}~\bibnamefont {{Das Sarma}}},\ }\href
  {https://doi.org/10.1103/physrevlett.121.245701} {\bibfield  {journal}
  {\bibinfo  {journal} {Phys. Rev. Lett.}\ }\textbf {\bibinfo {volume} {121}},\
  \bibinfo {pages} {245701} (\bibinfo {year} {2018})}\BibitemShut {NoStop}%
\bibitem [{\citenamefont {Khatami}\ \emph {et~al.}(2012)\citenamefont
  {Khatami}, \citenamefont {Rigol}, \citenamefont {Rela\~no},\ and\
  \citenamefont {Garc\'{\i}a-Garc\'{\i}a}}]{Khatami2012}%
  \BibitemOpen
  \bibfield  {author} {\bibinfo {author} {\bibfnamefont {E.}~\bibnamefont
  {Khatami}}, \bibinfo {author} {\bibfnamefont {M.}~\bibnamefont {Rigol}},
  \bibinfo {author} {\bibfnamefont {A.}~\bibnamefont {Rela\~no}},\ and\
  \bibinfo {author} {\bibfnamefont {A.~M.}\ \bibnamefont
  {Garc\'{\i}a-Garc\'{\i}a}},\ }\href
  {https://doi.org/10.1103/PhysRevE.85.050102} {\bibfield  {journal} {\bibinfo
  {journal} {Phys. Rev. E}\ }\textbf {\bibinfo {volume} {85}},\ \bibinfo
  {pages} {050102(R)} (\bibinfo {year} {2012})}\BibitemShut {NoStop}%
\bibitem [{\citenamefont {Singh}\ \emph {et~al.}(2017)\citenamefont {Singh},
  \citenamefont {Moessner},\ and\ \citenamefont {Roy}}]{Singh2017}%
  \BibitemOpen
  \bibfield  {author} {\bibinfo {author} {\bibfnamefont {R.}~\bibnamefont
  {Singh}}, \bibinfo {author} {\bibfnamefont {R.}~\bibnamefont {Moessner}},\
  and\ \bibinfo {author} {\bibfnamefont {D.}~\bibnamefont {Roy}},\ }\href
  {https://doi.org/10.1103/PhysRevB.95.094205} {\bibfield  {journal} {\bibinfo
  {journal} {Phys. Rev. B}\ }\textbf {\bibinfo {volume} {95}},\ \bibinfo
  {pages} {094205} (\bibinfo {year} {2017})}\BibitemShut {NoStop}%
\bibitem [{\citenamefont {De~Tomasi}(2019)}]{Giuseppe2019}%
  \BibitemOpen
  \bibfield  {author} {\bibinfo {author} {\bibfnamefont {G.}~\bibnamefont
  {De~Tomasi}},\ }\href {https://doi.org/10.1103/PhysRevB.99.054204} {\bibfield
   {journal} {\bibinfo  {journal} {Phys. Rev. B}\ }\textbf {\bibinfo {volume}
  {99}},\ \bibinfo {pages} {054204} (\bibinfo {year} {2019})}\BibitemShut
  {NoStop}%
\bibitem [{\citenamefont {Botzung}\ \emph {et~al.}(2019)\citenamefont
  {Botzung}, \citenamefont {Vodola}, \citenamefont {Naldesi}, \citenamefont
  {M\"uller}, \citenamefont {Ercolessi},\ and\ \citenamefont
  {Pupillo}}]{Thomas2019}%
  \BibitemOpen
  \bibfield  {author} {\bibinfo {author} {\bibfnamefont {T.}~\bibnamefont
  {Botzung}}, \bibinfo {author} {\bibfnamefont {D.}~\bibnamefont {Vodola}},
  \bibinfo {author} {\bibfnamefont {P.}~\bibnamefont {Naldesi}}, \bibinfo
  {author} {\bibfnamefont {M.}~\bibnamefont {M\"uller}}, \bibinfo {author}
  {\bibfnamefont {E.}~\bibnamefont {Ercolessi}},\ and\ \bibinfo {author}
  {\bibfnamefont {G.}~\bibnamefont {Pupillo}},\ }\href
  {https://doi.org/10.1103/PhysRevB.100.155136} {\bibfield  {journal} {\bibinfo
   {journal} {Phys. Rev. B}\ }\textbf {\bibinfo {volume} {100}},\ \bibinfo
  {pages} {155136} (\bibinfo {year} {2019})}\BibitemShut {NoStop}%
\bibitem [{\citenamefont {Yao}\ \emph {et~al.}(2014)\citenamefont {Yao},
  \citenamefont {Laumann}, \citenamefont {Gopalakrishnan}, \citenamefont
  {Knap}, \citenamefont {Müller}, \citenamefont {Demler},\ and\ \citenamefont
  {Lukin}}]{Yao2014}%
  \BibitemOpen
  \bibfield  {author} {\bibinfo {author} {\bibfnamefont {N.}~\bibnamefont
  {Yao}}, \bibinfo {author} {\bibfnamefont {C.~R.}\ \bibnamefont {Laumann}},
  \bibinfo {author} {\bibfnamefont {S.}~\bibnamefont {Gopalakrishnan}},
  \bibinfo {author} {\bibfnamefont {M.}~\bibnamefont {Knap}}, \bibinfo {author}
  {\bibfnamefont {M.}~\bibnamefont {Müller}}, \bibinfo {author} {\bibfnamefont
  {E.~A.}\ \bibnamefont {Demler}},\ and\ \bibinfo {author} {\bibfnamefont
  {M.~D.}\ \bibnamefont {Lukin}},\ }\href
  {https://doi.org/10.1103/physrevlett.113.243002} {\bibfield  {journal}
  {\bibinfo  {journal} {Phys. Rev. Lett.}\ }\textbf {\bibinfo {volume} {113}},\
  \bibinfo {pages} {243002} (\bibinfo {year} {2014})}\BibitemShut {NoStop}%
\bibitem [{\citenamefont {Burin}(2015)}]{Burin2015}%
  \BibitemOpen
  \bibfield  {author} {\bibinfo {author} {\bibfnamefont {A.~L.}\ \bibnamefont
  {Burin}},\ }\href {https://doi.org/10.1103/physrevb.92.104428} {\bibfield
  {journal} {\bibinfo  {journal} {Phys. Rev. B}\ }\textbf {\bibinfo {volume}
  {92}},\ \bibinfo {pages} {104428} (\bibinfo {year} {2015})}\BibitemShut
  {NoStop}%
\bibitem [{\citenamefont {Gutman}\ \emph {et~al.}(2016)\citenamefont {Gutman},
  \citenamefont {Protopopov}, \citenamefont {Burin}, \citenamefont {Gornyi},
  \citenamefont {Santos},\ and\ \citenamefont {Mirlin}}]{Gutman2016}%
  \BibitemOpen
  \bibfield  {author} {\bibinfo {author} {\bibfnamefont {D.~B.}\ \bibnamefont
  {Gutman}}, \bibinfo {author} {\bibfnamefont {I.~V.}\ \bibnamefont
  {Protopopov}}, \bibinfo {author} {\bibfnamefont {A.~L.}\ \bibnamefont
  {Burin}}, \bibinfo {author} {\bibfnamefont {I.~V.}\ \bibnamefont {Gornyi}},
  \bibinfo {author} {\bibfnamefont {R.~A.}\ \bibnamefont {Santos}},\ and\
  \bibinfo {author} {\bibfnamefont {A.~D.}\ \bibnamefont {Mirlin}},\ }\href
  {https://doi.org/10.1103/physrevb.93.245427} {\bibfield  {journal} {\bibinfo
  {journal} {Phys. Rev. B}\ }\textbf {\bibinfo {volume} {93}},\ \bibinfo
  {pages} {245427} (\bibinfo {year} {2016})}\BibitemShut {NoStop}%
\bibitem [{\citenamefont {Wu}\ and\ \citenamefont {{Das
  Sarma}}(2016)}]{Wu2016}%
  \BibitemOpen
  \bibfield  {author} {\bibinfo {author} {\bibfnamefont {Y.-L.}\ \bibnamefont
  {Wu}}\ and\ \bibinfo {author} {\bibfnamefont {S.}~\bibnamefont {{Das
  Sarma}}},\ }\href {https://doi.org/10.1103/physreva.93.022332} {\bibfield
  {journal} {\bibinfo  {journal} {Phys. Rev. A}\ }\textbf {\bibinfo {volume}
  {93}},\ \bibinfo {pages} {022332} (\bibinfo {year} {2016})}\BibitemShut
  {NoStop}%
\bibitem [{\citenamefont {Nandkishore}\ and\ \citenamefont
  {Sondhi}(2017)}]{Nandkishore_2017}%
  \BibitemOpen
  \bibfield  {author} {\bibinfo {author} {\bibfnamefont {R.~M.}\ \bibnamefont
  {Nandkishore}}\ and\ \bibinfo {author} {\bibfnamefont {S.~L.}\ \bibnamefont
  {Sondhi}},\ }\href {https://doi.org/10.1103/physrevx.7.041021} {\bibfield
  {journal} {\bibinfo  {journal} {Phys. Rev. X}\ }\textbf {\bibinfo {volume}
  {7}},\ \bibinfo {pages} {041021} (\bibinfo {year} {2017})}\BibitemShut
  {NoStop}%
\bibitem [{\citenamefont {Akhtar}\ \emph {et~al.}(2018)\citenamefont {Akhtar},
  \citenamefont {Nandkishore},\ and\ \citenamefont {Sondhi}}]{Akhtar_2018}%
  \BibitemOpen
  \bibfield  {author} {\bibinfo {author} {\bibfnamefont {A.~A.}\ \bibnamefont
  {Akhtar}}, \bibinfo {author} {\bibfnamefont {R.~M.}\ \bibnamefont
  {Nandkishore}},\ and\ \bibinfo {author} {\bibfnamefont {S.~L.}\ \bibnamefont
  {Sondhi}},\ }\href {https://doi.org/10.1103/physrevb.98.115109} {\bibfield
  {journal} {\bibinfo  {journal} {Phys. Rev. B}\ }\textbf {\bibinfo {volume}
  {98}},\ \bibinfo {pages} {115109} (\bibinfo {year} {2018})}\BibitemShut
  {NoStop}%
\bibitem [{\citenamefont {Roy}\ and\ \citenamefont {Logan}(2019)}]{Roy2019}%
  \BibitemOpen
  \bibfield  {author} {\bibinfo {author} {\bibfnamefont {S.}~\bibnamefont
  {Roy}}\ and\ \bibinfo {author} {\bibfnamefont {D.~E.}\ \bibnamefont
  {Logan}},\ }\href {https://doi.org/10.21468/SciPostPhys.7.4.042} {\bibfield
  {journal} {\bibinfo  {journal} {SciPost Phys.}\ }\textbf {\bibinfo {volume}
  {7}},\ \bibinfo {pages} {042} (\bibinfo {year} {2019})}\BibitemShut {NoStop}%
\bibitem [{\citenamefont {Schreiber}\ \emph {et~al.}(2015)\citenamefont
  {Schreiber}, \citenamefont {Hodgman}, \citenamefont {Bordia}, \citenamefont
  {Luschen}, \citenamefont {Fischer}, \citenamefont {Vosk}, \citenamefont
  {Altman}, \citenamefont {Schneider},\ and\ \citenamefont
  {Bloch}}]{Schreiber2015_Science}%
  \BibitemOpen
  \bibfield  {author} {\bibinfo {author} {\bibfnamefont {M.}~\bibnamefont
  {Schreiber}}, \bibinfo {author} {\bibfnamefont {S.~S.}\ \bibnamefont
  {Hodgman}}, \bibinfo {author} {\bibfnamefont {P.}~\bibnamefont {Bordia}},
  \bibinfo {author} {\bibfnamefont {H.~P.}\ \bibnamefont {Luschen}}, \bibinfo
  {author} {\bibfnamefont {M.~H.}\ \bibnamefont {Fischer}}, \bibinfo {author}
  {\bibfnamefont {R.}~\bibnamefont {Vosk}}, \bibinfo {author} {\bibfnamefont
  {E.}~\bibnamefont {Altman}}, \bibinfo {author} {\bibfnamefont
  {U.}~\bibnamefont {Schneider}},\ and\ \bibinfo {author} {\bibfnamefont
  {I.}~\bibnamefont {Bloch}},\ }\href {https://doi.org/10.1126/science.aaa7432}
  {\bibfield  {journal} {\bibinfo  {journal} {Science}\ }\textbf {\bibinfo
  {volume} {349}},\ \bibinfo {pages} {842} (\bibinfo {year}
  {2015})}\BibitemShut {NoStop}%
\bibitem [{\citenamefont {L\"{u}schen}\ \emph {et~al.}(2017)\citenamefont
  {L\"{u}schen}, \citenamefont {Bordia}, \citenamefont {Scherg}, \citenamefont
  {Alet}, \citenamefont {Altman}, \citenamefont {Schneider},\ and\
  \citenamefont {Bloch}}]{Lueschen2017_PRL}%
  \BibitemOpen
  \bibfield  {author} {\bibinfo {author} {\bibfnamefont {H.~P.}\ \bibnamefont
  {L\"{u}schen}}, \bibinfo {author} {\bibfnamefont {P.}~\bibnamefont {Bordia}},
  \bibinfo {author} {\bibfnamefont {S.}~\bibnamefont {Scherg}}, \bibinfo
  {author} {\bibfnamefont {F.}~\bibnamefont {Alet}}, \bibinfo {author}
  {\bibfnamefont {E.}~\bibnamefont {Altman}}, \bibinfo {author} {\bibfnamefont
  {U.}~\bibnamefont {Schneider}},\ and\ \bibinfo {author} {\bibfnamefont
  {I.}~\bibnamefont {Bloch}},\ }\href
  {https://doi.org/10.1103/physrevlett.119.260401} {\bibfield  {journal}
  {\bibinfo  {journal} {Phys. Rev. Lett.}\ }\textbf {\bibinfo {volume} {119}},\
  \bibinfo {pages} {260401} (\bibinfo {year} {2017})}\BibitemShut {NoStop}%
\bibitem [{\citenamefont {Lüschen}\ \emph {et~al.}(2018)\citenamefont
  {Lüschen}, \citenamefont {Scherg}, \citenamefont {Kohlert}, \citenamefont
  {Schreiber}, \citenamefont {Bordia}, \citenamefont {Li}, \citenamefont {{Das
  Sarma}},\ and\ \citenamefont {Bloch}}]{Lueschen2018_PRL}%
  \BibitemOpen
  \bibfield  {author} {\bibinfo {author} {\bibfnamefont {H.~P.}\ \bibnamefont
  {Lüschen}}, \bibinfo {author} {\bibfnamefont {S.}~\bibnamefont {Scherg}},
  \bibinfo {author} {\bibfnamefont {T.}~\bibnamefont {Kohlert}}, \bibinfo
  {author} {\bibfnamefont {M.}~\bibnamefont {Schreiber}}, \bibinfo {author}
  {\bibfnamefont {P.}~\bibnamefont {Bordia}}, \bibinfo {author} {\bibfnamefont
  {X.}~\bibnamefont {Li}}, \bibinfo {author} {\bibfnamefont {S.}~\bibnamefont
  {{Das Sarma}}},\ and\ \bibinfo {author} {\bibfnamefont {I.}~\bibnamefont
  {Bloch}},\ }\href {https://doi.org/10.1103/physrevlett.120.160404} {\bibfield
   {journal} {\bibinfo  {journal} {Phys. Rev. Lett.}\ }\textbf {\bibinfo
  {volume} {120}},\ \bibinfo {pages} {160404} (\bibinfo {year}
  {2018})}\BibitemShut {NoStop}%
\bibitem [{\citenamefont {Kohlert}\ \emph {et~al.}(2019)\citenamefont
  {Kohlert}, \citenamefont {Scherg}, \citenamefont {Li}, \citenamefont
  {L\"{u}schen}, \citenamefont {{Das Sarma}}, \citenamefont {Bloch},\ and\
  \citenamefont {Aidelsburger}}]{Kohlert2019_PRL}%
  \BibitemOpen
  \bibfield  {author} {\bibinfo {author} {\bibfnamefont {T.}~\bibnamefont
  {Kohlert}}, \bibinfo {author} {\bibfnamefont {S.}~\bibnamefont {Scherg}},
  \bibinfo {author} {\bibfnamefont {X.}~\bibnamefont {Li}}, \bibinfo {author}
  {\bibfnamefont {H.~P.}\ \bibnamefont {L\"{u}schen}}, \bibinfo {author}
  {\bibfnamefont {S.}~\bibnamefont {{Das Sarma}}}, \bibinfo {author}
  {\bibfnamefont {I.}~\bibnamefont {Bloch}},\ and\ \bibinfo {author}
  {\bibfnamefont {M.}~\bibnamefont {Aidelsburger}},\ }\href
  {https://doi.org/10.1103/physrevlett.122.170403} {\bibfield  {journal}
  {\bibinfo  {journal} {Phys. Rev. Lett.}\ }\textbf {\bibinfo {volume} {122}},\
  \bibinfo {pages} {170403} (\bibinfo {year} {2019})}\BibitemShut {NoStop}%
\bibitem [{Sup()}]{Supplement}%
  \BibitemOpen
  \href@noop {} {}\bibinfo {howpublished} {See Supplemental Material for
  Thermodynamic extrapolation, Level statistics, and Domain wall description in
  the strongly interacting limit with the additional reference
  ~\cite{Atlas2013}}\BibitemShut {NoStop}%
\bibitem [{\citenamefont {Nag}\ and\ \citenamefont {Garg}(2017)}]{Nag2017}%
  \BibitemOpen
  \bibfield  {author} {\bibinfo {author} {\bibfnamefont {S.}~\bibnamefont
  {Nag}}\ and\ \bibinfo {author} {\bibfnamefont {A.}~\bibnamefont {Garg}},\
  }\href {https://doi.org/10.1103/PhysRevB.96.060203} {\bibfield  {journal}
  {\bibinfo  {journal} {Phys. Rev. B}\ }\textbf {\bibinfo {volume} {96}},\
  \bibinfo {pages} {060203(R)} (\bibinfo {year} {2017})}\BibitemShut {NoStop}%
\bibitem [{\citenamefont {Bera}\ \emph {et~al.}(2015)\citenamefont {Bera},
  \citenamefont {Schomerus}, \citenamefont {Heidrich-Meisner},\ and\
  \citenamefont {Bardarson}}]{Bera2015}%
  \BibitemOpen
  \bibfield  {author} {\bibinfo {author} {\bibfnamefont {S.}~\bibnamefont
  {Bera}}, \bibinfo {author} {\bibfnamefont {H.}~\bibnamefont {Schomerus}},
  \bibinfo {author} {\bibfnamefont {F.}~\bibnamefont {Heidrich-Meisner}},\ and\
  \bibinfo {author} {\bibfnamefont {J.~H.}\ \bibnamefont {Bardarson}},\ }\href
  {https://doi.org/10.1103/physrevlett.115.046603} {\bibfield  {journal}
  {\bibinfo  {journal} {Phys. Rev. Lett.}\ }\textbf {\bibinfo {volume} {115}},\
  \bibinfo {pages} {046603} (\bibinfo {year} {2015})}\BibitemShut {NoStop}%
\bibitem [{\citenamefont {Nag}\ and\ \citenamefont
  {Garg}(2019)}]{Sabyasachi2019}%
  \BibitemOpen
  \bibfield  {author} {\bibinfo {author} {\bibfnamefont {S.}~\bibnamefont
  {Nag}}\ and\ \bibinfo {author} {\bibfnamefont {A.}~\bibnamefont {Garg}},\
  }\href {https://doi.org/10.1103/PhysRevB.99.224203} {\bibfield  {journal}
  {\bibinfo  {journal} {Phys. Rev. B}\ }\textbf {\bibinfo {volume} {99}},\
  \bibinfo {pages} {224203} (\bibinfo {year} {2019})}\BibitemShut {NoStop}%
\bibitem [{\citenamefont {Mastropietro}(2015)}]{Mastropietro2015}%
  \BibitemOpen
  \bibfield  {author} {\bibinfo {author} {\bibfnamefont {V.}~\bibnamefont
  {Mastropietro}},\ }\href {https://doi.org/10.1103/PhysRevLett.115.180401}
  {\bibfield  {journal} {\bibinfo  {journal} {Phys. Rev. Lett.}\ }\textbf
  {\bibinfo {volume} {115}},\ \bibinfo {pages} {180401} (\bibinfo {year}
  {2015})}\BibitemShut {NoStop}%
\bibitem [{\citenamefont {Vu}\ and\ \citenamefont {{Das
  Sarma}}(2020)}]{Vu2020}%
  \BibitemOpen
  \bibfield  {author} {\bibinfo {author} {\bibfnamefont {D.~D.}\ \bibnamefont
  {Vu}}\ and\ \bibinfo {author} {\bibfnamefont {S.}~\bibnamefont {{Das
  Sarma}}},\ }\href {https://doi.org/10.1103/physrevb.101.125113} {\bibfield
  {journal} {\bibinfo  {journal} {Phys. Rev. B}\ }\textbf {\bibinfo {volume}
  {101}},\ \bibinfo {pages} {125113} (\bibinfo {year} {2020})}\BibitemShut
  {NoStop}%
\bibitem [{\citenamefont {De~Tomasi}\ \emph {et~al.}(2019)\citenamefont
  {De~Tomasi}, \citenamefont {Hetterich}, \citenamefont {Sala},\ and\
  \citenamefont {Pollmann}}]{DeTomas2019}%
  \BibitemOpen
  \bibfield  {author} {\bibinfo {author} {\bibfnamefont {G.}~\bibnamefont
  {De~Tomasi}}, \bibinfo {author} {\bibfnamefont {D.}~\bibnamefont
  {Hetterich}}, \bibinfo {author} {\bibfnamefont {P.}~\bibnamefont {Sala}},\
  and\ \bibinfo {author} {\bibfnamefont {F.}~\bibnamefont {Pollmann}},\ }\href
  {https://doi.org/10.1103/PhysRevB.100.214313} {\bibfield  {journal} {\bibinfo
   {journal} {Phys. Rev. B}\ }\textbf {\bibinfo {volume} {100}},\ \bibinfo
  {pages} {214313} (\bibinfo {year} {2019})}\BibitemShut {NoStop}%
\bibitem [{\citenamefont {Langlett}\ and\ \citenamefont
  {Xu}(2021)}]{Langlett2021}%
  \BibitemOpen
  \bibfield  {author} {\bibinfo {author} {\bibfnamefont {C.~M.}\ \bibnamefont
  {Langlett}}\ and\ \bibinfo {author} {\bibfnamefont {S.}~\bibnamefont {Xu}},\
  }\href {https://doi.org/10.1103/PhysRevB.103.L220304} {\bibfield  {journal}
  {\bibinfo  {journal} {Phys. Rev. B}\ }\textbf {\bibinfo {volume} {103}},\
  \bibinfo {pages} {L220304} (\bibinfo {year} {2021})}\BibitemShut {NoStop}%
\bibitem [{Note1()}]{Note1}%
  \BibitemOpen
  \bibinfo {note} {\textcolor {blue}{In particular, the IPR at fixed $V$ as a
  function of $U$ should demonstrate similar values for $U=0$ and $U\to \infty
  $. We note that the domain wall argument is exact in the strong-$U$ limit and
  should be applicable non-perturbatively (in $V$). However, the symmetry is
  not quantitatively exact for small systems but should improve as the system
  size grows ~\cite {Supplement}}}\BibitemShut {NoStop}%
\bibitem [{\citenamefont {Sala}\ \emph {et~al.}(2020)\citenamefont {Sala},
  \citenamefont {Rakovszky}, \citenamefont {Verresen}, \citenamefont {Knap},\
  and\ \citenamefont {Pollmann}}]{Pablo2020}%
  \BibitemOpen
  \bibfield  {author} {\bibinfo {author} {\bibfnamefont {P.}~\bibnamefont
  {Sala}}, \bibinfo {author} {\bibfnamefont {T.}~\bibnamefont {Rakovszky}},
  \bibinfo {author} {\bibfnamefont {R.}~\bibnamefont {Verresen}}, \bibinfo
  {author} {\bibfnamefont {M.}~\bibnamefont {Knap}},\ and\ \bibinfo {author}
  {\bibfnamefont {F.}~\bibnamefont {Pollmann}},\ }\href
  {https://doi.org/10.1103/PhysRevX.10.011047} {\bibfield  {journal} {\bibinfo
  {journal} {Phys. Rev. X}\ }\textbf {\bibinfo {volume} {10}},\ \bibinfo
  {pages} {011047} (\bibinfo {year} {2020})}\BibitemShut {NoStop}%
\bibitem [{\citenamefont {Khemani}\ \emph {et~al.}(2020)\citenamefont
  {Khemani}, \citenamefont {Hermele},\ and\ \citenamefont
  {Nandkishore}}]{Vedika2020}%
  \BibitemOpen
  \bibfield  {author} {\bibinfo {author} {\bibfnamefont {V.}~\bibnamefont
  {Khemani}}, \bibinfo {author} {\bibfnamefont {M.}~\bibnamefont {Hermele}},\
  and\ \bibinfo {author} {\bibfnamefont {R.}~\bibnamefont {Nandkishore}},\
  }\href {https://doi.org/10.1103/PhysRevB.101.174204} {\bibfield  {journal}
  {\bibinfo  {journal} {Phys. Rev. B}\ }\textbf {\bibinfo {volume} {101}},\
  \bibinfo {pages} {174204} (\bibinfo {year} {2020})}\BibitemShut {NoStop}%
\bibitem [{\citenamefont {Atas}\ \emph {et~al.}(2013)\citenamefont {Atas},
  \citenamefont {Bogomolny}, \citenamefont {Giraud},\ and\ \citenamefont
  {Roux}}]{Atlas2013}%
  \BibitemOpen
  \bibfield  {author} {\bibinfo {author} {\bibfnamefont {Y.~Y.}\ \bibnamefont
  {Atas}}, \bibinfo {author} {\bibfnamefont {E.}~\bibnamefont {Bogomolny}},
  \bibinfo {author} {\bibfnamefont {O.}~\bibnamefont {Giraud}},\ and\ \bibinfo
  {author} {\bibfnamefont {G.}~\bibnamefont {Roux}},\ }\href
  {https://doi.org/10.1103/PhysRevLett.110.084101} {\bibfield  {journal}
  {\bibinfo  {journal} {Phys. Rev. Lett.}\ }\textbf {\bibinfo {volume} {110}},\
  \bibinfo {pages} {084101} (\bibinfo {year} {2013})}\BibitemShut {NoStop}%
\end{thebibliography}%


\begin{thebibliography}{1}%
\makeatletter
\providecommand \@ifxundefined [1]{%
 \@ifx{#1\undefined}
}%
\providecommand \@ifnum [1]{%
 \ifnum #1\expandafter \@firstoftwo
 \else \expandafter \@secondoftwo
 \fi
}%
\providecommand \@ifx [1]{%
 \ifx #1\expandafter \@firstoftwo
 \else \expandafter \@secondoftwo
 \fi
}%
\providecommand \natexlab [1]{#1}%
\providecommand \enquote  [1]{``#1''}%
\providecommand \bibnamefont  [1]{#1}%
\providecommand \bibfnamefont [1]{#1}%
\providecommand \citenamefont [1]{#1}%
\providecommand \href@noop [0]{\@secondoftwo}%
\providecommand \href [0]{\begingroup \@sanitize@url \@href}%
\providecommand \@href[1]{\@@startlink{#1}\@@href}%
\providecommand \@@href[1]{\endgroup#1\@@endlink}%
\providecommand \@sanitize@url [0]{\catcode `\\12\catcode `\$12\catcode
  `\&12\catcode `\#12\catcode `\^12\catcode `\_12\catcode `\%12\relax}%
\providecommand \@@startlink[1]{}%
\providecommand \@@endlink[0]{}%
\providecommand \url  [0]{\begingroup\@sanitize@url \@url }%
\providecommand \@url [1]{\endgroup\@href {#1}{\urlprefix }}%
\providecommand \urlprefix  [0]{URL }%
\providecommand \Eprint [0]{\href }%
\providecommand \doibase [0]{https://doi.org/}%
\providecommand \selectlanguage [0]{\@gobble}%
\providecommand \bibinfo  [0]{\@secondoftwo}%
\providecommand \bibfield  [0]{\@secondoftwo}%
\providecommand \translation [1]{[#1]}%
\providecommand \BibitemOpen [0]{}%
\providecommand \bibitemStop [0]{}%
\providecommand \bibitemNoStop [0]{.\EOS\space}%
\providecommand \EOS [0]{\spacefactor3000\relax}%
\providecommand \BibitemShut  [1]{\csname bibitem#1\endcsname}%
\let\auto@bib@innerbib\@empty
\bibitem [{\citenamefont {Atas}\ \emph {et~al.}(2013)\citenamefont {Atas},
  \citenamefont {Bogomolny}, \citenamefont {Giraud},\ and\ \citenamefont
  {Roux}}]{Atlas2013}%
  \BibitemOpen
  \bibfield  {author} {\bibinfo {author} {\bibfnamefont {Y.~Y.}\ \bibnamefont
  {Atas}}, \bibinfo {author} {\bibfnamefont {E.}~\bibnamefont {Bogomolny}},
  \bibinfo {author} {\bibfnamefont {O.}~\bibnamefont {Giraud}},\ and\ \bibinfo
  {author} {\bibfnamefont {G.}~\bibnamefont {Roux}},\ }\href
  {https://doi.org/10.1103/PhysRevLett.110.084101} {\bibfield  {journal}
  {\bibinfo  {journal} {Phys. Rev. Lett.}\ }\textbf {\bibinfo {volume} {110}},\
  \bibinfo {pages} {084101} (\bibinfo {year} {2013})}\BibitemShut {NoStop}%
\end{thebibliography}%

\end{document}


\global\long\def\id{\mathbbm{1}}
\global\long\def\ui{\mathbbm{i}}
\global\long\def\ud{\mathrm{d}}

\title{Supplemental Material for ``Fermionic many-body localization for random and quasiperiodic systems 
in the presence of short- and long-range interactions''}

\author{DinhDuy Vu}
\affiliation{Condensed Matter Theory Center and Joint Quantum Institute, University of Maryland, College Park, Maryland 20742, USA}
\author{Ke Huang}
\affiliation{Department of Physics, City University of Hong Kong, Kowloon, Hong Kong SAR}
\author{Xiao Li}
\affiliation{Department of Physics, City University of Hong Kong, Kowloon, Hong Kong SAR}
\affiliation{City University of Hong Kong Shenzhen Research Institute, Shenzhen 518057, Guangdong, China}
\author{S. Das Sarma}
\affiliation{Condensed Matter Theory Center and Joint Quantum Institute, University of Maryland, College Park, Maryland 20742, USA}

\setcounter{equation}{0} 
\setcounter{figure}{0}
\setcounter{table}{0} 
\renewcommand{\theparagraph}{\bf}
\renewcommand{\thefigure}{S\arabic{figure}}
\renewcommand{\theequation}{S\arabic{equation}}
\renewcommand\labelenumi{(\theenumi)}

\maketitle

In this Supplemental Material, we present additional details on the thermodynamic extrapolation of mean IPR results, the level statistics results, and the domain wall theoretical description in the strong interaction limit. 

\section{Thermodynamic extrapolation and scaling characteristics}

\subsection{The extrapolation algorithm}

\begin{figure}[t]
  \includegraphics[width=0.48\textwidth]{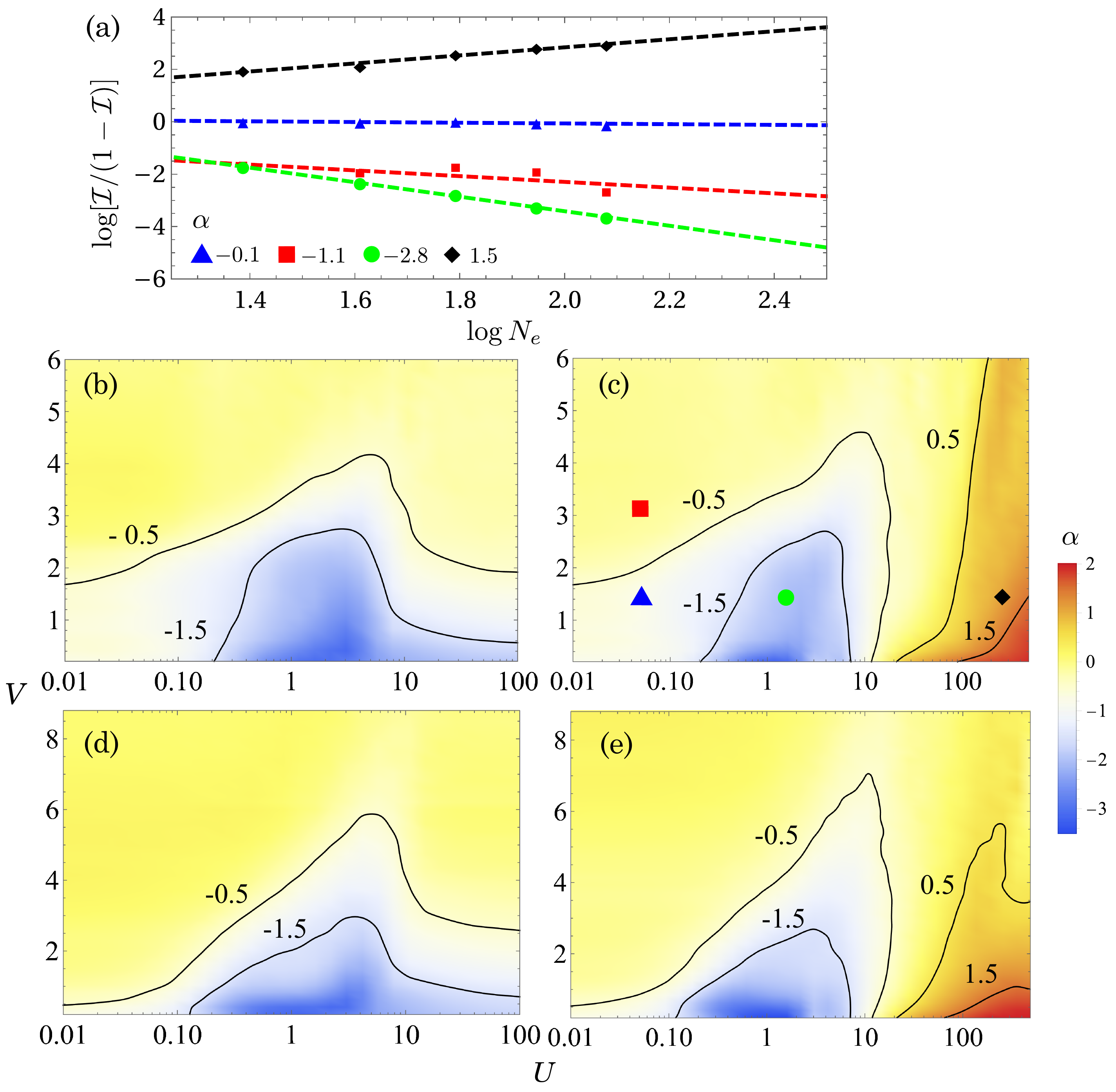}
  \caption{(a) Scaling with respect to the electron number in the AA, long-range case. The coordinates of the points, also shown in (c), are: $U=0.05, V=1.4$ (blue triangle), $U=0.05, V = 3.1$ (red square), $U=1.5, V=1.4$ (green circle), and $U=260, V = 1.4$ (black diamond). (b-d) Extracted scaling exponent for AA-SR, AA-LR, A-SR, A-LR respectively. Isolines corresponding to $\alpha=-1.5, -0.5, 0.5, 1.5$ are provided.  \label{Fig7}}
\end{figure}

In this section, we provide the algorithm for the extrapolation used in the main text. We first perform the exact diagonalization on half-filled finite systems with $L=8,10,\dots,16$. The Hamiltonian matrices are constructed in the Fock space, i.e. each basis is an occupancy configuration. The number of basis (the dimension of the Hilbert space) is thus $d=N!/[(N/2)!]^2$. The matrix is then diagonalized numerically, giving the full spectrum.  We note that all the Hilbert space is included in the exact diagonalization. After obtaining the spectrum average IPR $\expval{I}$ at each finite size, we perform a linear fit in the logarithmic scale based on the ansatz
   \begin{equation}
       \log \left( \frac{\expval{\mathcal{I}}}{1-\expval{\mathcal{I}}} \right) = \alpha\log N_e + \beta.
   \end{equation}
In Fig.~\ref{Fig7}(a), we present the scaling of a representative point for each scaling zones defined in the main text, all of which agree with the power law ansatz. The extracted exponents are shown in Figs.~\ref{Fig7}(c-e) as a compliment to Fig. (2) of the main text. However, the logarithmic scale does not allow extrapolation to the thermodynamic limit $N_e\to\infty$, so we apply the second data fitting defined piecewise with respect to the exponent $\alpha$
   \begin{align}
        & \alpha > 1  && \frac{1-\expval{\mathcal{I}}}{\expval{\mathcal{I}}} = \beta^{-1}\left( \frac{1}{N_e} \right)^\alpha + \frac{A_1}{N_e} + A_0 \nonumber\\
        & 1\ge \alpha >0  && \frac{1-\expval{\mathcal{I}}}{\expval{\mathcal{I}}} = \frac{A_1}{N_e} + A_0 \nonumber\\
        & 0\ge \alpha > -1  && \frac{\expval{\mathcal{I}}}{1-\expval{\mathcal{I}}}= \frac{A_1}{N_e} + A_0  \nonumber\\
        & -1\ge \alpha  && \frac{\expval{\mathcal{I}}}{1-\expval{\mathcal{I}}} = \beta\left( \frac{1}{N_e} \right)^{-\alpha} + \frac{A_1}{N_e} + A_0.
   \end{align}
This algorithm, by definition, makes all the transitions smooth. As a result, we cannot determine whether the phase transition involves a divergent derivative or a continuous crossover. However, away from the boundary, we expect the extrapolated algorithm to provide reasonable predictions. 
The exponent $\alpha$ is thus an effective exponent, which may indicate a crossover behavior.

\subsection{Additional details on the scaling analysis}

We now elaborate on the relevant scaling behavior of the strong LR interaction MBL phase. As mentioned in the main text, hybridization can only occur within bands comprised of translation-related configurations. The single-particle hopping can only shift one electron at a time, thus the possible intra-band coupling must be high-order perturbations. For demonstration, we exemplify one perturbative channel. The true hopping amplitude should be multiplied by an extra factor $\sim \mathcal{O}(N_e)$ counting the number of similar-order channels. Beginning at an initial Fock basis state $\ket{\psi_\text{int}}=\{x_1,\dots,x_{N_e}\}$, we apply one electron hopping at a time so that the intermediate states are $\ket{\psi_n} = \{x_1+1,\dots,x_n+1,x_{n+1},\dots,x_{N_e}\}$ ($n=1,\dots,N_e-1$) until a global translation is obtained and the final state is  $\ket{\psi_\text{fin}}=\{x_1,\dots,x_{N_e}\} + 1$. We express this process through the unperturbed Hamiltonian (disorder potential and interaction energy) and the perturbation (electron hopping) as follows
\begin{align}
  & H_0=\begin{pmatrix}
   K_\text{int} & 0 &  & \\
    0 & \eta_1 U + K_1 & & \\
    & 0 & \ddots & 0 & \\
    & & 0 & \eta_{N_e-1} U +  K_{N_e-1} & 0 \\
    & & & 0 & K_\text{fin} \\
  \end{pmatrix}, \nonumber \\
  & H_1 =\begin{pmatrix}
   0 & t &  & \\
    t & 0 & & \\
    & t & \ddots & t & \\
    & & t & 0 & t \\
    & & & t & 0 \\
  \end{pmatrix}. 
\end{align}
Here, 
the potential energy is  $K_n = \expval{V}{\psi_n}$ and the many-body interaction energy is expressed through $\eta_n = (U_n - U_\text{int})/U \sim \mathcal{O}(1)$. We note that $U_\text{fin}=U_\text{int}$ by the translational symmetry. 
We assume the limit $U\gg K \gg t$. The effective coupling between $\ket{\psi_\text{int}}$ and $\ket{\psi_\text{fin}}$ can be obtained through the series with two projections $P=\ket{\psi_\text{int}}\bra{\psi_\text{int}} + \ket{\psi_\text{fin}}\bra{\psi_\text{fin}}$ and $Q=\ket{\psi_{n}}\bra{\psi_{n}}$
\begin{equation}
\begin{split}
    &H_\text{eff}  = PH_0P + PH_1P  \\
     & +PH_1Q\frac{1}{E_0-QH_0Q}\sum_k \qty[\frac{QH_1Q-E+E_0}{{E_0-QH_0Q}}]^k QH_1P. \notag
\end{split}
\end{equation}
Due to the vanishing first order perturbation $E-E_0 \sim \mathcal{O}(t^\epsilon) \ll t$ with $\epsilon > 1$. We also have $E_0-QH_0Q\approx U \mathbf{1}$ (assuming $\eta_n=1$ for simplicity). The lowest-order non-zero coupling, corresponding to $k=N_e-2$, is then 
\begin{equation}
    t_\text{eff} \approx t(t/U)^{N_e-1}.
\end{equation}
There are other ways to construct the intermediate states but the order of perturbation (or the number of intermediate states) is the most important element.

On the other hand, the typical band splitting  $\Delta K = K_\text{fin}-K_\text{int} \sim V N_e^{1/2}$ by the central limit theorem (we also verified this statement numerically). Suppose at $t=0$, we have the pure state $\ket{\psi_{t=0}} = \ket{\psi_\text{int}}$, the corresponding IPR is obviously unity as the sites can have electron occupancy of either zero or unity. Upon turning on the hopping, we have a weakly mixed state $\ket{\psi_{t=1}} = \sqrt{1-\delta^2}\ket{\psi_\text{int}} + \delta \ket{\psi_\text{fin}}$ (other long-range couplings are subleading) with $\delta = t_\text{eff}/|\Delta K|$ obtained through the first-order perturbation theory. As a result, the occupied sites now have the electron occupancy $\approx 1-\delta^2$ while some initially unoccupied sites may have the electron occupancy $\propto \delta^2$. Then $1-\expval{\mathcal{I}} \propto \delta^2 \sim U^{-2N_e}$. Here, we ignore the polynomial scaling of $|\Delta K|$ as it is much slower than the exponential decay of $t_\text{eff}$. We note that the expression we derive is clearly not a power law as used in the scaling analysis and the extracted $\alpha$ is only meaningful for a small variation range of the scaling parameter; nevertheless, $\alpha$ must be positive, already distinguishing this phase with the disorder-driven localized phase. The exponent $\alpha$ should be considered as an effective exponent characterizing the various interaction regimes of MBL physics.

\subsection{The many-body IPR in the $U\to0$ limit}

It is well known that in the noninteracting ($U\to0$) limit Anderson localization in a 1D system takes place for an infinitesimal amount of disorder. 
However, one may be confused by Fig.~1(c) and 1(d) of the main text, where the many-body IPR $\expval{\mathcal{I}}$ with $U=0.01$ seems to rise slowly from zero as $V$ is increased. 
This is in contrast to the AA model results shown in Fig.~1(a) and 1(b), where the many-body IPR $\expval{\mathcal{I}}$ with $U=0.01$ rises much more rapidly from zero at the single-particle localization transition point $V=2$. 

\begin{figure}[b]
  \includegraphics[width=0.48\textwidth]{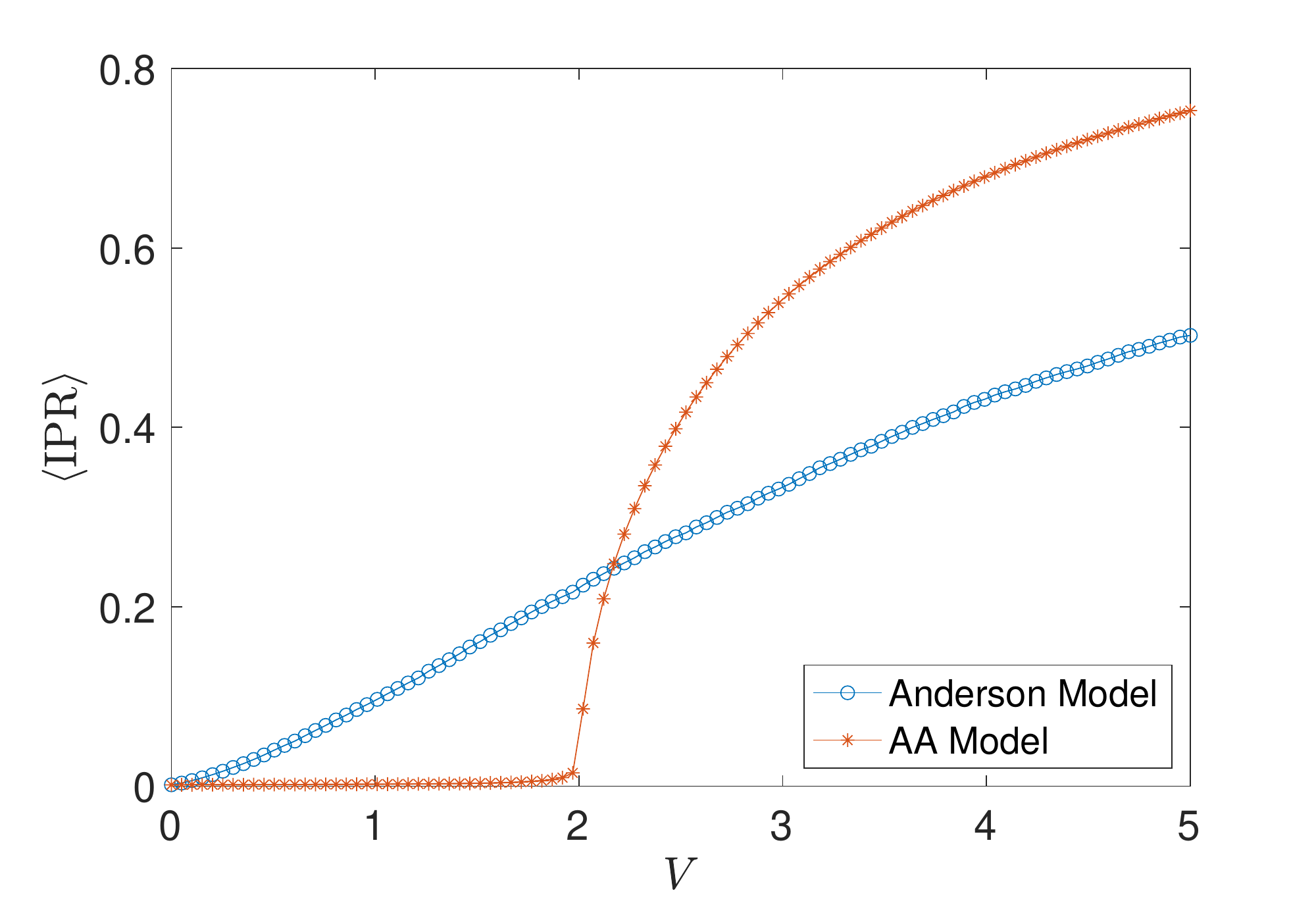}
  \caption{ \label{Fig9}
  The mean single-particle IPR $\expval{I_s}$ in the Anderson and AA model in a system with $L=10^3$ sites. 
  }
\end{figure}

As we now argue, the observed differences between the two models are not just finite size effects, but are intrinsic to the two models, as shown in Fig.~\ref{Fig9}. 
Before we discuss the details of this figure, we first note that in the noninteracting limit ($U=0$) the average many-body IPR [as we defined in Eq.~(3) in the main text] is equal to the average single-particle IPR in the thermodynamic limit $L\to\infty$. 

To prove this statement, we note that in the non-interacting case, the eigenstates of the system can be described by a unitary matrix
\begin{align}
	w_{ji}=\braket{j}{\psi_i}, 
\end{align}
where $\ket{j}$ denotes the $j$th lattice site and $\ket{\psi_i}$ denotes the $i$th single-particle eigenstate. 
Consequently, the mean single-particle IPR $\expval{I_s}$ is defined as
\begin{align}
	\expval{I_s} =\frac1L\sum_{i,j}\abs{w_{ij}}^4.
\end{align}
Meanwhile, for a system with $L$ sites containing $N<L$ particles, the average many-body IPR is defined as
\begin{align}
	\expval{\mathcal{I}(L, N)} =\frac{X-1}{1-\nu},
\end{align}
where we introduced for convenience
\begin{align}
	X=\frac1{N\binom{L}{N}}\sum_{i,\lambda}\ev{n_i}{\psi_\lambda}^2. 
\end{align}
In the above equation, $\binom{L}{N} \equiv \frac{L!}{N!(L-N)!}$ is the binomial coefficient. 
Noting that in the noninteracting limit the many-body eigenstates are product states of single-particle orbitals and the fact that 
\begin{align}
	\ev{n_i}{\psi_\lambda}=\sum_{j\in \Omega_\lambda}\abs{w_{ij}}^2,
\end{align}
$X$ can be rewritten as
\begin{widetext}
\begin{align}
	X&=\frac{1}{N\binom{L}{N}}\sum_{\abs{\Omega}=N}\sum_i\Biggl[\sum_{j\in\Omega}\abs{w_{ij}}^2\Biggr]^2
	=\frac{1}{N\binom{L}{N}}\sum_{\abs{\Omega}=N}\sum_i\sum_{j,k\in\Omega}\abs{w_{ij}}^2\abs{w_{ik}}^2 \notag \\
	&=\frac{1}{N}\dfrac{\binom{L-1}{N-1}}{\binom{L}{N}}\sum_i\sum_{j}\abs{w_{ij}}^4+\frac{1}{N}\frac{\binom{L-2}{N-2}}{\binom{L}{N}}\sum_i\sum_{j\neq k}\abs{w_{ij}}^2\abs{w_{ik}}^2 \\
	&=\sum_i\sum_{j}\dfrac{\abs{w_{ij}}^4}{L}+\frac{1}{L}\frac{N-1}{L-1}\sum_i\sum_{j\neq k}\abs{\psi_{ij}}^2\abs{w_{ik}}^2
	=\frac{L-N}{L-1}\expval{I_s}
  +\frac{1}{L}\frac{N-1}{L-1}\sum_i\Biggl[\sum_{j}\abs{w_{ij}}^2\Biggr]^2 
	=\frac{L-N}{L-1}\expval{I_s}+\frac{N-1}{L-1}. \notag
\end{align}
\end{widetext}
In the above derivation $\sum_{\abs{\Omega}=N}$ denotes summation over all $N$-element subset $\Omega$ of $\{1,2,...,L\}$. 
In addition, the third equality is obtained by counting how many $\Omega$ contains a given pair of $(j,k)$. The last equality uses the unitarity of $\psi_{ij}$.
Consequently, we find that 
\begin{align}
	\expval{\mathcal{I}(L, N)}
  &=\frac1{1-\nu}\qty(X-1) \notag\\
	&=\frac{L}{L-N}\frac{L-N}{L-1}\expval{I_s}+\frac{L}{L-N}\qty(\frac{N-1}{L-1}-\frac{N}{L}) \notag \\
	&=\frac{\expval{I_s}-1/L}{1-1/L}. 
\end{align}
Therefore, we have 
\begin{align}
  \lim_{L \to \infty} \expval{\mathcal{I}(L, N)} = \expval{I_s}. 
\end{align}
The above result shows that when $U=0$ the average many-body IPR is equal to the average single-particle IPR in the thermodynamic limit. 

With this relation established, we can now use the average single-particle IPR to understand the average many-body IPR in the $U\to0$ limit in Fig.~1 in the main text. 
Such a result is shown in Fig.~\ref{Fig9}. 
We can find that the average IPR in the Anderson model increases very slowly from zero as $V$ is turned on. 
In contrast, the average IPR in the AA model increases very sharply from zero at the single-particle localization transition point $V=2$. 
Such a difference can help us understand why the average many-body IPR in the Anderson model does not increase very rapidly from zero when $U = 0.01$. 
    
\section{Level statistics}
\begin{figure}
\includegraphics[width=0.48\textwidth]{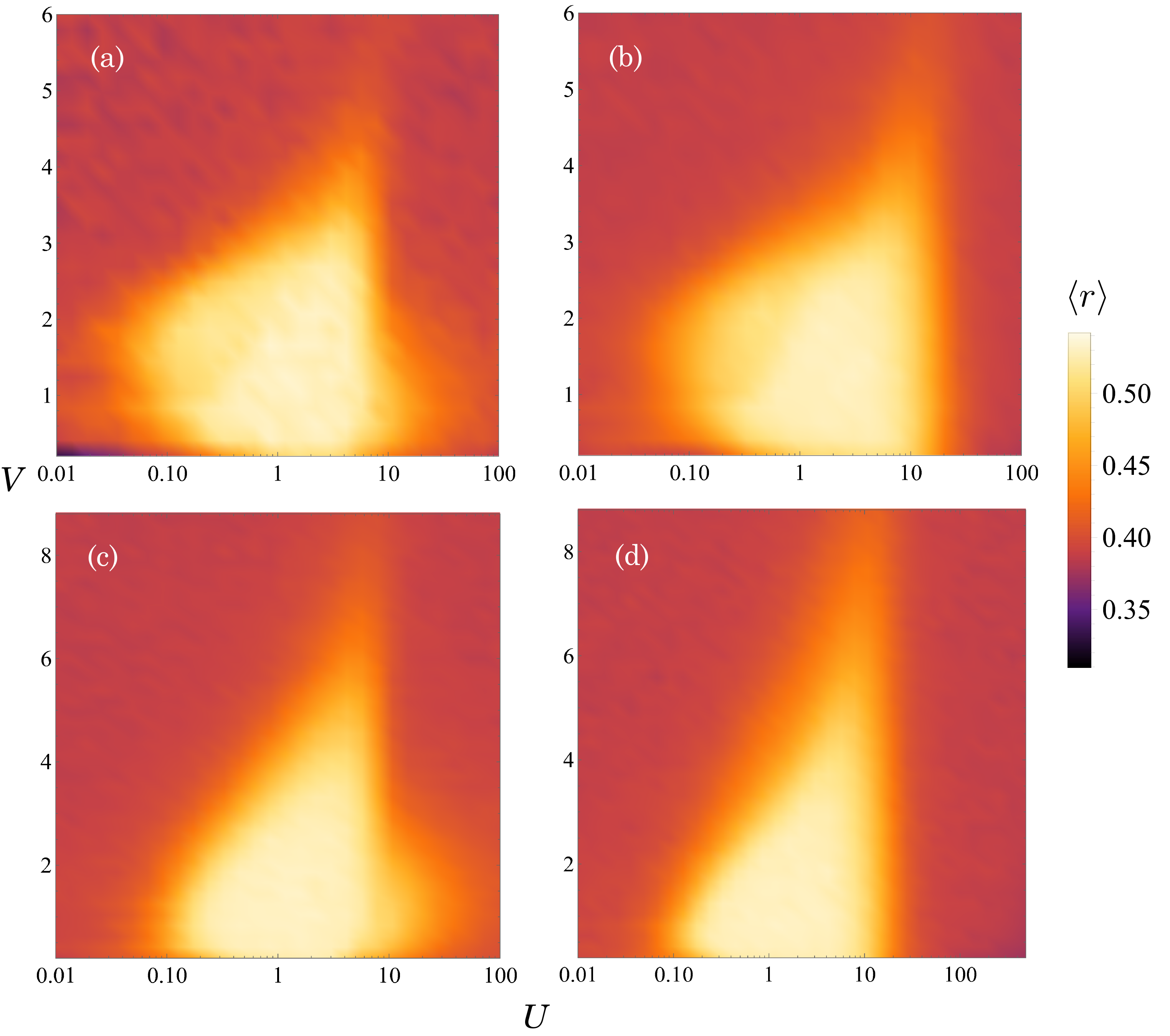}
\caption{\label{Fig4}
Level uniformity $\expval{r}$ in an $L=16$ system, corresponding to Fig. 1 in the main text. (a)-(d) show the results for AA model (SR), AA model (LR), A model (SR), and A model (LR), respectively. 
}
\end{figure}

Complementary to the IPR results presented in the main text, the level statistics also indicates the localized or extended degrees of the system. 
Ideally, the system with all extended eigenstates should have a level spacing distribution close to the Gaussian Orthogonal Ensemble (GOE) distribution, featuring a strong level repulsion. 
By contrast, a localized system with uncorrelated eigenvalues should follow the Poisson distribution. 
The statistics can be studied using the normalized level spacing $s$ or level uniformity $r$, defined as 
\begin{equation}
s_n = \frac{E_n - E_{n-1}}{\langle s \rangle}, \quad 
r_n = \frac{\text{min}(s_n,s_{n-1})}{\text{max}(s_n,s_{n-1})}, 
\end{equation}
where $\expval{s}$ is the average level spacing over all eigenstates. 
The GOE distributions as defined through $s$ and $r$ are approximately given by 
\begin{equation}
f_{G}(s) = \frac{\pi}{2} se^{-\pi s^2/4}, \quad 
f_{G}(r) = \frac{27}{4}\frac{r+r^2}{\left(1+r+r^2\right)^{5/2}}. 
\end{equation}
A more accurate expression can be found in Ref.~\cite{Atlas2013} The corresponding expressions for the Poisson distributions are
\begin{equation}
    f_{P}(s) = e^{-s},\quad 
    f_{P}(r) = \frac{1}{(1+r)^2}. 
\end{equation}
We provide in Figs.~\ref{Fig4} the level uniformity statistics for different models in a system with $L=16$ sites. 
Specifically, $\expval{r} \approx 0.53$ for GOE distribution and $\expval{r}\approx 0.38$ for Poisson distribution. Therefore, $\expval{r}$ is high (low) for extended (localized) system. 
It is easy to see the agreement between the region with GOE distribution and the super-reciprocal extended region with scaling exponent $\alpha<-1$. 
  	
\section{Domain wall description in the strongly interacting limit}
\begin{figure}
	\includegraphics[width=0.48\textwidth]{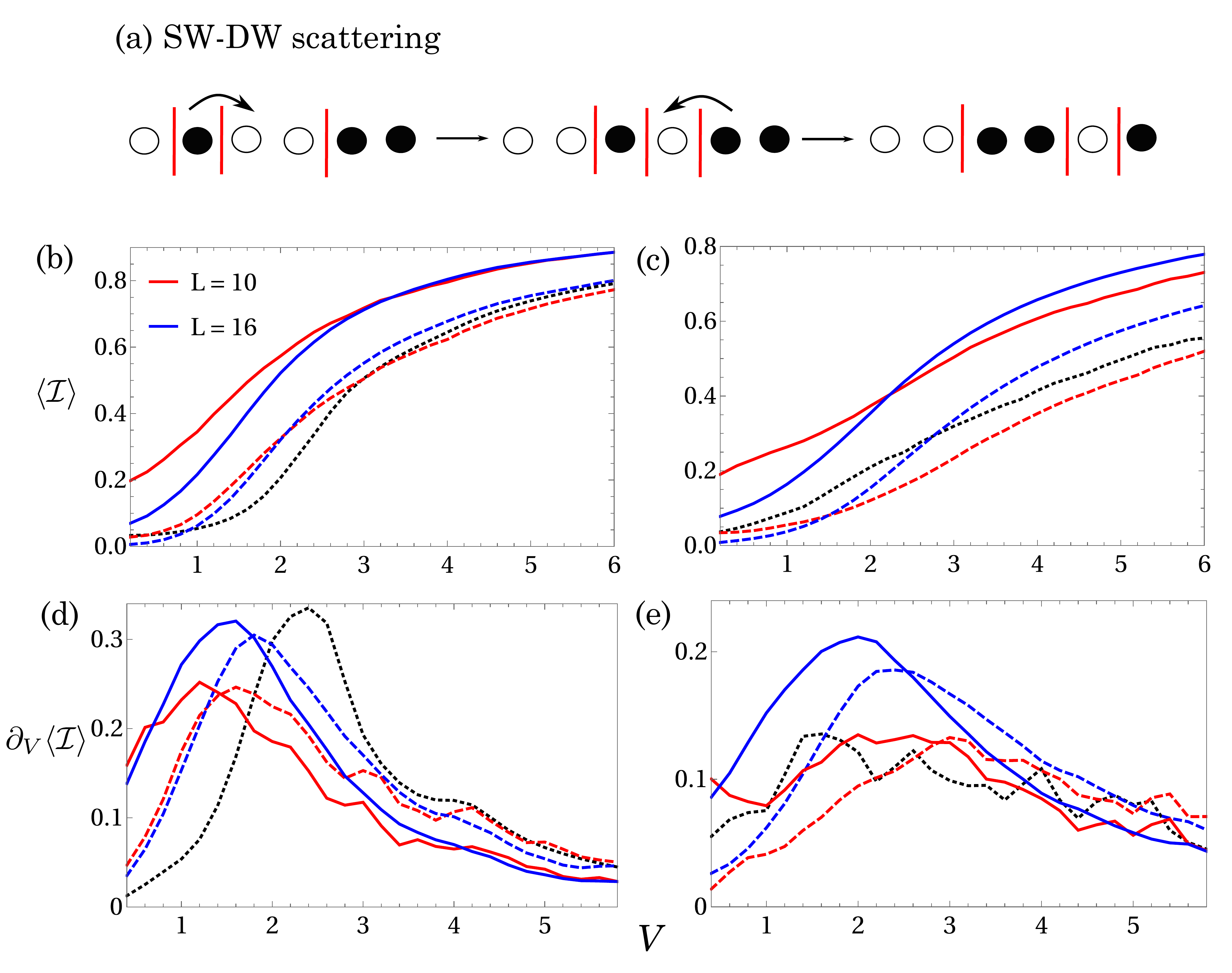}
	\caption{(a) SW-DW scattering that can shift the otherwise immobile SW. (b-e) Original infinite-$U$ mean IPR (solid lines) and with single domain walls removed (dash lines) for the AA-SR and A-SR respectively. The same color represents the same lattice size. The black dotted line is the IPR of non-interacting model at $L=16$ for reference. (c-d) First derivatives of mean IPR with respect to $V$ correspond to (a-b). \label{Fig8}}
\end{figure} 

In this section, we elaborate on the domain wall picture used in the main text to explain the symmetry between $U\to0$ and $U\to\infty$ limits in the SR interaction case. 
From Fig.~3 in the main text, it is clear that for a single domain wall (SW) a single electron hopping process (equivalently spin flip in the Heisenberg spin chain) necessarily creates two more walls, and thus is forbidden. 
In contrast, the same operator applied to a double wall (DW) only moves its position without changing the number of domain walls. 
In addition, each time a DW moves by one lattice site, a potential energy difference of $V_{i+1}-V_i$ is involved, similar to the hopping of a single electron across the disordered lattice. 
Therefore, the ballistic propagation of DWs leads to the $V$-dependence observed in $\expval{\mathcal{I}}$ at strong interaction and the symmetry with the zero-$U$ limit. 

However, certain effects are unique in the strong SR interaction limit: 
	(a) the contribution of immobile SWs to the density profile, overshadowing the signature of itinerant DWs,  
	(b) a DW can scatter with an SW [see Fig.~\ref{Fig8}(a)], and 
	(c) each band has a different number of domain walls. 
	We anticipate that these factors make the finite-size effects so severe that the $\expval{\mathcal{I}}$s for the SR interaction at weak and strong $U$ in our phase diagrams are visibly different.
As we now argue, these factors are only meaningful for small lattices and will vanish in sufficiently large systems, recovering the exact symmetry we hypothesize. 
Every time a propagating DW scatters with an SW, the latter is shifted by two lattice sites. 
As a result, if a band contains both itinerant SWs and DWs, such scattering events eventually delocalize all domain walls. 
Therefore, effect (a) is actually alleviated by (b), except for configurations containing only SWs, such configurations are statistically unlikely in the thermodynamic limit from the following argument.
We consider a band having $n_D$ domain walls and relax the condition of fixed filling so that these domain walls can be placed on any sites. 
If there are no DWs, each SW must occupy at least 3 sites (no neighboring domain walls to the left and right), leading to the number of configurations in the leading order $\mathcal{N}_1 = LC_{n_D-1}^{L/3}$ where $C_m^n$ is the binomial coefficient. 
Compared to the total number of configurations $\mathcal{N}_2 = C_{n_D}^{L}$, the ratio $\mathcal{N}_1/\mathcal{N}_2 < (1+n_D/L)^{n_D}/n_D \to 0$ in the limit $n,L \to \infty$. 
We expect this conclusion to hold even when the fixed filling constraint is imposed.
   
As shown in Fig.~\ref{Fig8}(b-e), there exist a quantitatively visible difference between the strong- and weak-$U$ limits that is less severe as the system size grows (comparing solid lines representing $L=10$ and 16). From the first derivative of IPR (i.e., $\partial_V\expval{\mathcal{I}}$), the strongly interacting system does inherit some features of the noninteracting model, particularly the prominent peak in $\partial_V \expval{\mathcal{I}}$ for the AA model at $V=2$. The agreement also improves with the increasing system size. To show that the finite size effect mostly stems from the immobile SWs manifesting , we remove them from the density profile (but still keep them in the Hamiltonian diagonalization). 
Specifically, each eigenstate is the superposition of basis states having the same number of DWs and SWs, since the hopping term cannot change their numbers but only move their positions. 
We thus keep the mixing coefficients, but only count for the DWs in each basis state to obtain the density profile of the superimposed eigenstate. 
This treatment excludes eigenstates comprised entirely from SWs. 
The results in Fig.~\ref{Fig8}(b-e) as dashed lines show a significant improvement. 
Specifically, the treated mean IPR now has a weaker finite-size effect and its value is closer to that of the noninteracting counterpart in both $\expval{\mathcal{I}}$ and its first derivative with respect to $V$.

\bibliographystyle{apsrev4-2}
\bibliography{AA_MBL_v1}